\documentclass[aps, prl, twocolumn, superscriptaddress, 10pt]{revtex4-2}  
\usepackage{amssymb, amsmath, amsthm}
\usepackage{xcolor}
\usepackage{graphicx}
\usepackage{hyperref}
\usepackage{comment}
\usepackage{mathtools} 
\usepackage{tikz}
\usetikzlibrary{shapes.geometric}

\newcommand{\ket}[1]{| #1 \rangle}
\newcommand{\bra}[1]{\langle #1 |}

\newcommand{\mytitle}{Eightfold way to dark states in SU($3$) cold gases with two-body losses}

\graphicspath{{./Figures/}}

\begin{document}
  
\title{\mytitle}      

\author{Lorenzo Rosso}
\affiliation{Universit\'e Paris-Saclay, CNRS, LPTMS, 91405 Orsay, France}

\author{Leonardo Mazza}
\affiliation{Universit\'e Paris-Saclay, CNRS, LPTMS, 91405 Orsay, France}

\author{Alberto Biella}
\email{alberto.biella@unitn.it}
\affiliation{INO-CNR BEC Center and Dipartimento di Fisica, Universit\`a di Trento, 38123 Povo, Italy}
\affiliation{Universit\'e Paris-Saclay, CNRS, LPTMS, 91405 Orsay, France}

\begin{abstract}
We study the quantum dynamics of a one-dimensional  SU($3$)-symmetric system of cold atoms in the presence of two-body losses.
We exploit the representation theory of SU($3$), the so-called eightfold way, as a scheme to organize the dark states of the dissipative dynamics in terms of generalized Dicke states and show how they are dynamically approached, both in the weakly- and and strongly-interacting and dissipative regimes.
Our results are relevant for a wide class of alkaline-earth(-like) gases experiments, paving the way to the dissipative preparation and exploitation of generalized Dicke states.
\end{abstract}

\maketitle

\textbf{Introduction --} 
Ultracold atomic gases represent a clean and flexible playground to study quantum many-body physics, at equilibrium or in dynamical settings~\cite{LangenRev_2015,Gross2017,Schafer2020}.
Cold-atom experiments usually feature a high degree of control over system parameters and  allow for an almost perfect decoupling from the surrounding environment. 
However, despite the tremendous experimental progresses, a perfect isolation has never been reached, for instance because of particle losses, causing energy relaxation and decoherence phenomena~\cite{Zurek_2003}.  
On one hand, this fact introduces a typical timescale determining for how long a system can be regarded as {\itshape closed}. 
On the other hand, on a longer timescale, the interplay between the coherent unitary evolution and the coupling to the environment can lead to a non-trivial dynamics and to stationary states featuring strong quantum correlations~\cite{Syassen_2008,GarciaRipoll_2009,Kantian_2009,
Letscher_2017} and critical behaviors~\cite{Diehl_2008,Lee_2011,Jin_2016,Morsch_2018}. 

In general, this latter situation can be achieved through an active control of the environment and of its coupling to the system, via the so-called reservoir engineering~\cite{Verstraete_2009}; however, in some situations, the dissipative processes that naturally occur in the system can also be responsible for entangled stationary states: this is the situation that we want to study in this letter~\cite{Sponselee_2018,Rosso_2021}.
Since in these systems decoherence is mainly due to particle losses, developing a theoretical framework to describe this open-system dynamics and the emergence of eventual correlated quantum states represents a huge theoretical challenge that attracted an increasing attention in the recent years~\cite{Kordas_2015, Johnson_2017, Schemmer_2018, Bouchoule_2020b, Bouchoule_2020SCP, Ashida_2020,Bouchoule_2021, Bouchoule_2021PRA,Nakagawa_2020,Nakagawa_2021}.
In particular, two-body losses induced by inelastic atomic collisions  in correlated quantum gases have been observed experimentally and investigated theoretically in bosonic~\cite{Syassen_2008,GarciaRipoll_2009,Durr_2009,Tomita_2017,Tomita_2019,Rossini_2020,Rosso_2021bis,Scarlatella_2021,Secli_2022} 
and fermionic gases~\cite{Zhu_2014,Yan_2013,FossFeig_2012,Kazuki_2019,Kazuki_2021,Rosso_2021}.

In this work we consider the paradigmatic case of alkaline-earth-like gases in optical lattices, experimentally realized with ytterbium~\cite{Scazza_2014,Pagano_2014,Franchi_2017,Bouganne_2017,Sponselee_2018}, which are subject to two-body losses due to inelastic two-body collisions in the metastable state $\prescript{3}{}{P}_{0}$.
The (almost) perfect decoupling between the nuclear spin $I$ and the electronic angular momentum $J$ (ensured by the fact that $J=0$ for the atomic states involved in the dynamics) implies that the relevant scattering processes are independent of $I$.
As a result, this class of systems has an emergent SU($N$) spin symmetry (with $N=2I+1$) whose dynamics is governed by a SU($N$)-symmetric Fermi-Hubbard model describing alkaline-earth-like atoms in an optical lattice~\cite{Gorshkov_2010,Cazalilla_2014}.
In the two-spin case ($N=2$) the dissipative dynamics conserves the total spin and the system exhibits stationary states that are a mixture of highly-entangled wavefunctions with a Dicke-like spin component~\cite{FossFeig_2012,Rosso_2021}, 
which could be exploited for various quantum-technology purposes. 
The impact of two-body losses for $N>2$ has not been theoretically addressed at present, despite the availability of experimental data obtained in this regime~\cite{Sponselee_2018}.

In this letter we study the quantum dynamics of an interacting SU($3$)-symmetric one-dimensional fermionic gas in the presence of two-body losses.  
We show that the dark states of the dynamics can be organized via the representation theory of this group, the so-called {\itshape eightfold way}~\cite{GeorgiBook}. 
This elegant classification allows us to characterise a family of stationary states using the notion of generalised Dicke states~\cite{Hartmann_2016} describing the spin degrees of freedom of the gas.
Next, we discuss the system dynamics highlighting how the generalized Dicke-like states represent the unique attractor of the dynamics both in the weakly-dissipative and weakly-interacting limit as well as in the strongly-dissipative and strongly-interacting quantum Zeno regime.
Finally, we draw our conclusions and discuss future perspectives.

\textbf{The model --} 
Introducing the fermionic operators $\hat c_{j,\mu}^{(\dagger)}$ (with $j$ and $\mu$ labelling the lattice site and the spin, respectively), which satisfy canonical anticommutation relations, the SU($N$)-symmetric Fermi-Hubbard Hamiltonian reads:
\begin{equation}
\label{sunham}
 \hat H = - J \sum_{j,\mu} \left(\hat c_{j, \mu}^\dagger \hat c_{j+1, \mu}+ {\rm H.c.}\right) + U \sum_{j,\mu<\mu'}\hat n_{j, \mu} \hat n_{j, \mu'}.
\end{equation}
Here, $J$ is the hopping amplitude, $U$ is the spin-independent interaction strength and $\hat n_{j,\mu} = \hat c^\dagger_{j,\mu} \hat c_{j, \mu}$ is the spin-resolved on-site lattice-density operator. 
The spin index can assume $N$ values that in the following will be labelled with capital letters in progressive order ($\mu=A,B,C,\dots$).
The Hamiltonian~\eqref{sunham} is invariant under global SU($N$) rotations in spin space.
As a consequence, the unitary dynamics conserves the expectation value of the $N(N-1)/2$ SU($2$) pseudo-spin algebra generators defined in each subspace (here labelled by $\mu\mu'$ with $\mu<\mu'$) as
\begin{equation}
\label{gensun}
\hat\Lambda^{\alpha}_{\mu\mu'} =\frac{1}{2}\sum_{j}
\begin{pmatrix}
\hat{c}_{j,\mu}^\dagger,\hat{c}_{j,\mu'}^\dagger  
\end{pmatrix}
\sigma^\alpha 
\begin{pmatrix}
\hat{c}_{j,\mu} \\
\hat{c}_{j,\mu'} 
\end{pmatrix},
\ \ \  \alpha=0,x,y,z
\end{equation}
where $\{\sigma^{\alpha}|\alpha=x,y,z\}$ are the Pauli matrices and \mbox{$\sigma^{0}=\mathbb{I}_{2}$}.

The presence of local two-body losses is accounted for  by the jump operators
\begin{equation}
\label{family_jump}
\hat L_{j,\mu\mu'} = \sqrt{\gamma}  \ \hat c_{j, \mu} \hat c_{j, \mu'},
\end{equation} 
with $j=1,\cdots,L$ and $\mu<\mu'$ and $\gamma$ being the dissipation rate.
The dynamics of the full density matrix $\rho(t)$ is described by a Lindblad master equation:
\begin{equation}
 \dot \rho(t) = - \frac{i}{\hbar} \left[ \hat H, \rho(t) \right] + \sum_{j,\mu<\mu'} \mathsf{D}_{j,\mu\mu'}[\rho(t)],
 \label{Eq:MEQ}
\end{equation}
with \mbox{$\mathsf{D}_{j,\mu\mu'}[\rho(t)]= \hat L_{j,\mu\mu'} \rho(t) \hat L_{j,\mu\mu'}^\dagger  - \frac12\{ \hat L_{j,\mu\mu'}^\dagger \hat L_{j,\mu\mu'}, \rho(t) \}$}.

The main difference with respect to the $N=2$ case is that the spin components defined in Eq.~\eqref{gensun} are not conserved quantities of the full dissipative dynamics: the breaking of these conservation laws is due to the presence of several spin sectors involved in the dynamics.
Thus, in terms of symmetries, the study of the $N=3$ case can be considered representative for all the $N>2$ models, which therefore will not be explicitly considered.

\textbf{Equations of motion and dark states --}  
Let us now focus on the population dynamics and define the total number of atoms $\hat N = \sum_{\mu}\hat N_{\mu}$, where $\hat N_{\mu}=\sum_{j}\hat n_{j,\mu}$ is the spin-resolved population.
In what follows we will use the notation $O(t)\doteqdot\langle \hat O \rangle_t\doteqdot\text{tr}[\rho(t) \hat O]$. 
The spin-resolved populations obey the following equation~\cite{SuppMat}
\begin{equation}
\dot N_{\mu}(t) = - \gamma \sum_j \sum_{\mu'\neq\mu} \Big\langle  \hat n_{j,\mu} \hat n_{j,\mu'} \Big\rangle_t.
 \label{Eq:N:Diss:1}
\end{equation}

First, we will present a construction allowing us to map out all the possible dark states of the dissipative dynamics factorizing spin and charge degrees of freedom. Such states are not affected by the dissipative dynamics and any statistical mixture of them is stationary with respect to the master equation~\eqref{Eq:MEQ}. Next, we will study the system dynamics showing how the system evolves, because of dissipation, towards such a dark subspace.
We consider the class of states where orbital and spin degrees of freedom factorize,
$
\ket{\Psi_{\rm dark}} = \ket{\Psi_{\rm orb}} \otimes \ket{\Psi_{\rm spin}}.
$
If $\ket{\Psi_{\rm orb}}$ is constructed as a Slater determinant of a set of appropriate orbital modes, i.e. the eigenstates of the hopping Hamiltonian in Eq.~\eqref{sunham}, the state is assured to commute with the Hamiltonian and never to have a double spatial occupation, so that no particle can leak out of it. 
Since the full many-body wave function $\ket{\Psi_{\rm dark}}$ of the system must be fully-antisymmetric, and one such $\ket{\Psi_{\rm orb}}$ is fully-antisymmetric, the spin wavefunction $\ket{\Psi_{\rm spin}}$ must be fully-symmetric.
In order to understand the properties of these states, we make use of group theory.

\begin{figure}[t]
 \includegraphics{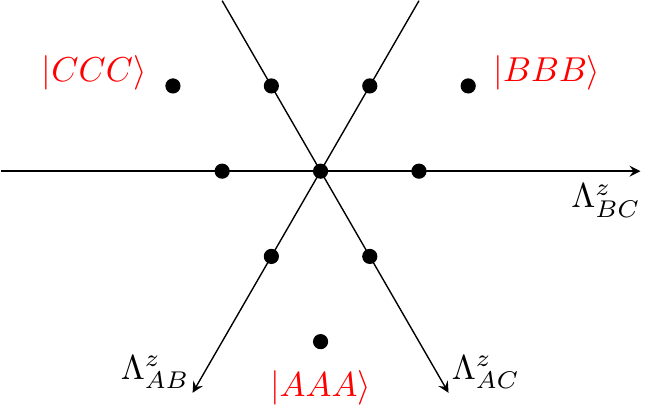}
 \caption{The eightfold way in a dark state. Triangular irreducible representation with labels $(3,0)$, which is composed of $10$ states. The three arrows allow to identify each state through the quantum numbers $\Lambda^z_{\mu\mu'}$, where $\mu, \mu'$ take values in the three components of the gas, $A$, $B$ and $C$. Note that only two of them are linearly independent.}
 \label{Fig:Eightfold:1}
\end{figure}

The irreducible representations of SU(3) are labeled by two integers $(p,q)$~\cite{GeorgiBook}; according to group theory, the fully-symmetric SU(3) states correspond to the representations with labels $(p,0)$ and the states belonging to it can be arranged in the shape of a triangle turned upside-down with edge length $p+1$, see Fig.~\ref{Fig:Eightfold:1} for an example with $p=3$. 
The number of particles accomodated in the representation is $N = p$; 
The dimension of a representation $(p,0)$ is $(p+1)(p+2)/2$, and each state is uniquely determined by the values of $\Lambda^z_{\mu \mu'}$.
In the case of the figure we have the ten fully-symmetric states of $N=3$ particles.
At the three vertices of the triangle we always find the fully-polarized states, in this case $\ket{AAA}$, $\ket{BBB}$ and $\ket{CCC}$.
The other states are obtained by repeated application of the spin-ladder operators $\hat \Lambda^\pm_{\mu\mu'}$. If we want, for instance, to construct all the states that stand on the top edge of the triangle, from left to right we need to  apply the operator $\hat \Lambda^+_{BC}$, that raises the value of $\Lambda_{BC}^z$ by one, starting from $\ket{CCC}$.

These states are \textit{generalised Dicke states}~\cite{Hartmann_2016} since they are fully-symmetric with respect to the exchange of two particles generalizing the symmetry properties of the stationary states of the SU(2) lossy dynamics identified in Ref.~\cite{Rosso_2021}.
Given two spin sectors $\mu,\mu'$, such states satisfy the relation~\cite{SuppMat}
\begin{equation}
\label{dickecondition}
\frac{\langle \hat S^2_{\mu\mu'}\rangle}{\hbar^2} = \left \langle \frac{\hat N_{\mu\mu' }}{2} \left(\frac{\hat N_{\mu\mu'}}{2} + 1 \right)\right \rangle,
\end{equation}
where $\hat S^{\alpha}_{\mu\mu'}= \hbar \hat\Lambda^{\alpha}_{\mu\mu'}$ for $\alpha=x,y,z$ and $\hat N_{\mu\mu'}=\hat N_{\mu} + \hat N_{\mu'}= 2 \hat \Lambda^{0}_{\mu\mu'}$.
Conversely, Eq.~\eqref{dickecondition} can be satisfied only by the generalized Dicke states. This can be explicitly seen by considering the irreducible representations of the SU($3$) group with $q\neq0$. These representations of the group are not fully-symmetric and, together with the $q=0$ case, cover all the possible spin states that can be constructed within SU($3$). By explicit construction of such states it is easy to see that for any $q\neq0$ we get 
$
\langle \hat S^2_{\mu\mu'}\rangle/\hbar^2 < \left \langle \hat N_{\mu\mu' }/2 (\hat N_{\mu\mu'}/2 + 1 )\right \rangle.
$

While via the eightfold way we constructed explicitly the dark states for $N=3$, our reasoning is general and generalized Dicke states are dark states of the master equation~\eqref{Eq:MEQ} for any $N$ and regardless of the specific values of the system parameters.

\textbf{Dynamics --} 
While it is true that such states surely are stationary states of the dynamics it is not trivial to show that they are unique. 
Indeed, our analysis focused on states where the spin and orbital part of the wavefunctions factorize while we can not exclude a priori that non-factorizable dark states exhist.

To corroborate this scenario, we will make use of Eq.~\eqref{dickecondition} certifying that the system has flown to a mixture of generalised Dicke states.
In what follows we will consider two paradigmatic regimes: (i) the weakly-dissipative and weakly-interacting regime and (ii) the strongly-dissipative and strongly-interacting limit.  

\textit{Weak dissipation and weak interactions--} 
We start by studying the regime of weak dissipation and weak interactions $\hbar \gamma,U\ll J$.
In this limit we can write the evolution of the spin-resolved densities as~\cite{SuppMat}
\begin{equation}
\label{gausssys_2}
 \dot n_{\mu}(t) = \gamma\sum_{\mu'\neq\mu}
 \vec{s}^{ {\ \mathsf T}}_{\mu\mu'} \ \mathsf{G} \ \vec{s}_{\mu\mu'},
\end{equation}
where we defined the four-component vector $\vec{s}_{\mu\mu'}=(s^{0}_{\mu\mu'},s^{x}_{\mu\mu'}/\hbar,s^{y}_{\mu\mu'}/\hbar,s^{z}_{\mu\mu'}/\hbar)$ with $s^{\alpha}_{\mu\mu'} (t)=\langle\hat S^{\alpha}_{\mu\mu'}\rangle_{t}/L$, $s^{0}_{\mu\mu'}=\langle\hat \Lambda^{0}_{\mu\mu'}\rangle_{t}/L$, $n_{\mu}=\langle \hat N_{\mu}\rangle/L$ and
$
\mathsf{G}={\rm diag}(-1,1,1,1)
$
being the relativistic Minkowsky tensor.

The fact that the time-derivative of spin-resolved populations is related to the Minkowski scalar product of a 4-component vector suggests some suggestive analogies with the theory of special relativity.
The structure of Eq.~\eqref{gausssys_2} highlights indeed some of the symmetries of the problem as the internal rotations of the SU($2$) pseudospins (indicating that the physics does not have a preferred direction in the internal space) and the analogs of the Lorentz boosts (which allow for the exchange between populations and coherences). Furthermore, the analogy with the Minkowski tensor, suggests an effective representation of the dynamics in a population-spin diagram, where the dynamics is constrained within an effective light cone, that we dubbed \textit{Dicke cone}.

Let us start by briefly reviewing the $N=2$ case.
In this case we just have two spin sectors labelled as $\mu=A,B$. Therefore, to determine the fixed points, we ask $\dot n_{A}=\dot n_{B}=0$.
From Eq.\eqref{gausssys_2} we get the following stationarity condition
\begin{equation}
\label{dickesu2}
\vec{s}^{ {\ \mathsf T}}_{AB} \ \mathsf{G} \ \vec{s}_{AB}=0
\ \Rightarrow \ 
s_{AB}=\frac\hbar2 n_{AB},
\end{equation}
where $s_{AB}=\sqrt{(s^{x}_{AB})^{2}+(s^{y}_{AB})^{2}+(s^{z}_{AB})^{2}}$.
The condition~\eqref{dickesu2} holds both for Dicke states ($N=2$) and generalized Dicke states ($N>2$)
\footnote{Taking the thermodynamic limit of Eq.~\eqref{dickecondition} we get 
$$
\lim_{L\to\infty}\frac{\langle\hat S^{2}_{AB}\rangle}{L^{2}} = \frac{\hbar^{2}}{4} \frac{\langle\hat N_{AB}^{2}\rangle}{L^{2}},
$$
which gives the relation~\eqref{dickesu2}.}
and defines the boundary of the {\it Dicke cone} within which the dynamics must take place because of the physical requirement $s_{AB}\leq\hbar \ n_{AB}/2$.
As a result, the system dynamics can be effectively visualized in a two-dimensional parameter space spanned by the variables $s_{AB}$ and $n_{AB}$ constrained to the Dicke cone.
Finally, since the $s_{AB}$ is a constant of motion for the $N=2$ case
$s_{AB}(t) = s_{AB}(0)$ and thus the dynamics must take place on the line defined by the initial value of the spin.
In the $t\to\infty$ limit, the boundary of the light cone are touched (i.e. $n_{AB}=2s_{AB}/\hbar$) and the system reaches a stable stationary state.
The $N=2$ case has been discussed extensively in Ref.~\cite{Rosso_2021}. The conservation of the total spin, even in the presence of dissipative events, plays a crucial role in constraining the system dynamics. 
Indeed, given the initial conditions, it allows to be predictive about the final density of the system. 
Starting from the $N=2$ case we want now to explore the $N>2$ case where the dynamics does not conserve the spin. 
 
Let us consider the $N=3$ case where the internal states are labelled as $\mu=A,B,C$.
In this case the spin components are no longer conserved and in general  
$s_{\mu\mu'}(t) \neq s_{\mu\mu'}(0)$.
\begin{figure}[t]
\includegraphics[width=0.7\columnwidth]{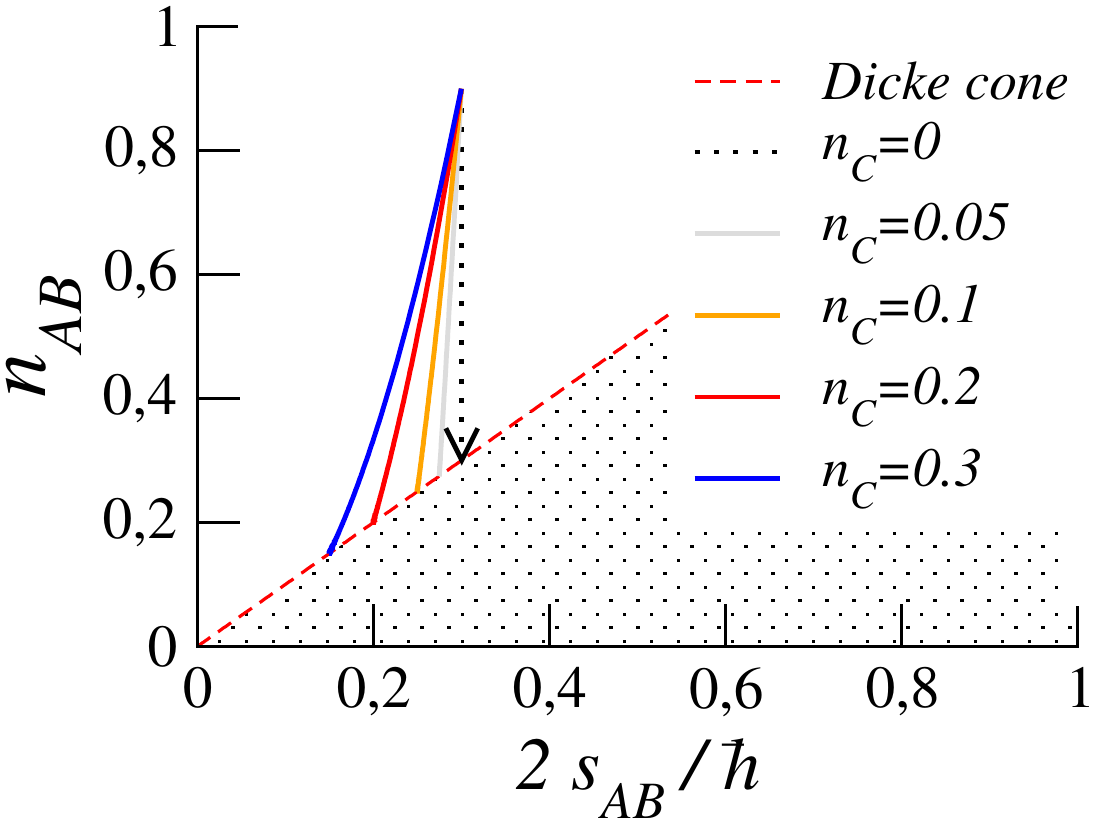}
\caption{SU($3$) dynamics in the $n_{AB}-s_{AB}$ plane. In the $n_{C}=0$ case the evolution must follow vertical lines defined by the initial value of $s_{AB}$. 
When $n_{C}>0$ the spin conservation does not hold. The dynamics escapes the vertical line defined by $s_{AB}(0)$ and deviates progressively towards $s_{AB}=0$ getting steady when $n_{AB}=2s_{AB}$. 
Here we set $n_{A}=0.5, n_{B}=0.4$, $(s_{AB}^{x}(0),s_{AB}^{y}(0),s_{AB}^{z}(0)) = (0.1,0.1,0.05)$ so that $2s_{AB}(0)=0.3$.}
\label{Fig:SU3_nc_v2}
\end{figure}
In Fig.~\ref{Fig:SU3_nc_v2} we show the dynamics of spin and number of particles in the $AB$ subspace for a generic initial condition.
When $n_{C}=0$ the dynamics is spin conserving $s_{AB}(t) = s_{AB}(0)$ and the system dynamics follows vertical lines. 
Even if an additional internal state is now available, there are no physical processes that populate it. As a result, in this limit the system behaves effectively as in the $N=2$ case.
For $n_{C}>0$ the spin in the $AB$ subspace is no longer conserved but  gets shrinked. 
The trajectory in the $n_{AB}-s_{AB}$ plane deviates on the left of the $s_{AB}(0)$ line and evolves until the boundary of the Dicke cone is approached.

\begin{figure}[t!]
\includegraphics[width=0.7\columnwidth]{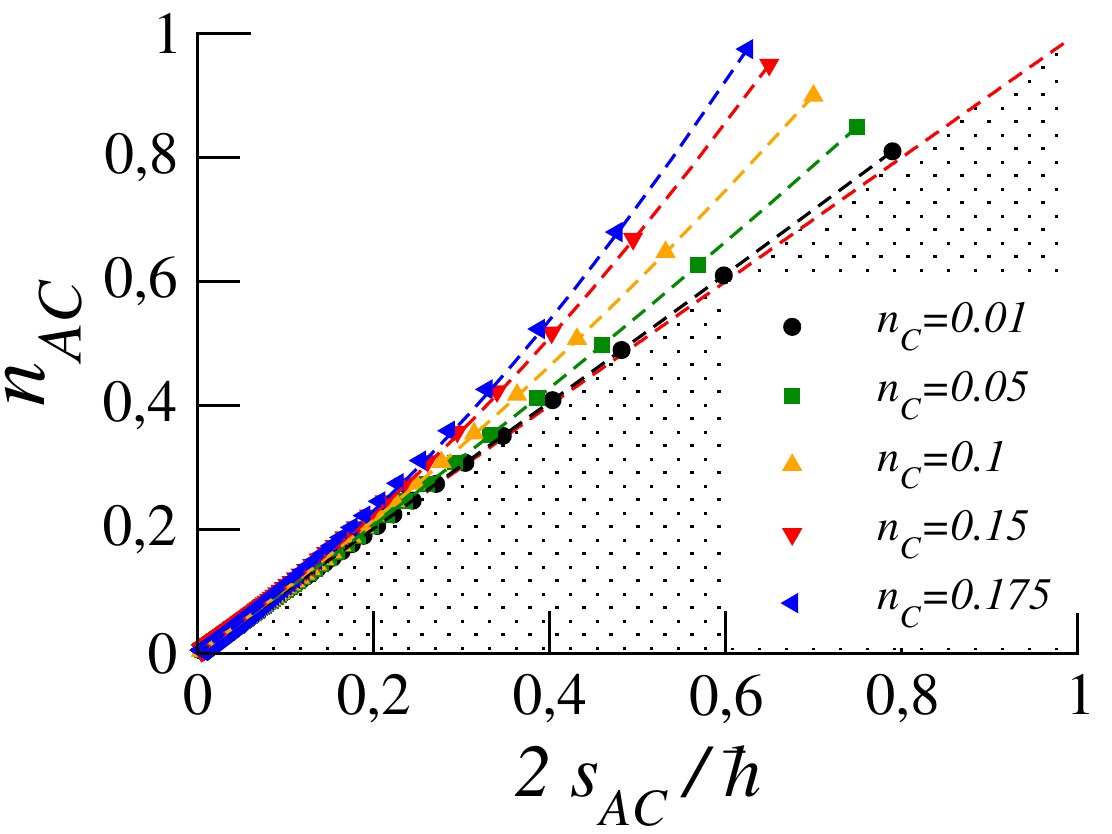}
\includegraphics[width=0.7\columnwidth]{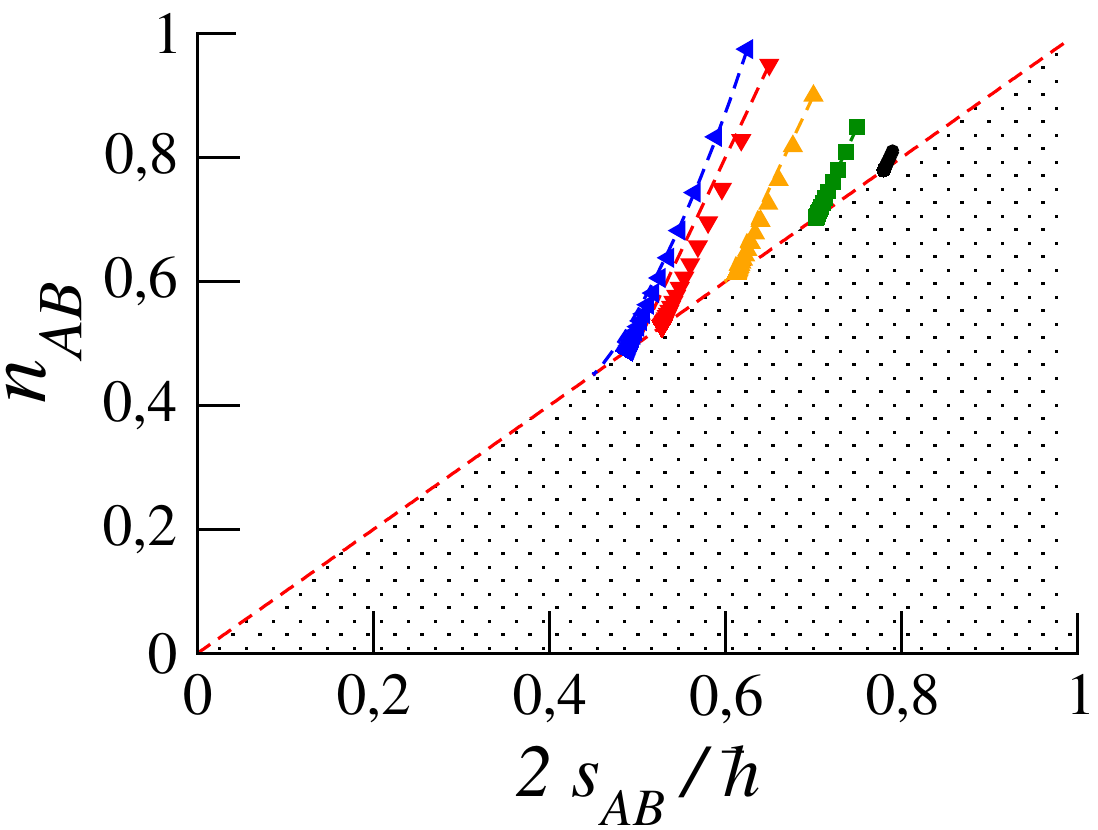}
\caption{SU($3$) dynamics in the weakly dissipative limit. 
Top panel: we set $n_{A}(0)=n_{B}(0)=0.8$ and $n_{C}(0)$ is varied. 
Bottom panel: we set $n_{A}(0)=0.8$ and $n_{B}(0)=n_{C}(0)$ is varied. 
In both the cases the numerics (filled symbols) shows a good agreement with the predictions (dashed lines) of Eq.~\eqref{exactCsmall} (for the top panel) and Eq.~\eqref{su3limB} (for the bottom panel), even beyond the limit $n_{C}(0)\ll 1$.
In all the panels $s^{x,y}_{\mu\mu'}=0, \forall \mu<\mu'$.}
\label{Fig:SU3_case12}
\end{figure}

We now propose a perturbative solution of the SU($3$) dynamics for different initial conditions considering the experimentally-relevant situation where  $s^{x,y}_{\mu\mu'}=0, \forall \mu<\mu'$.
We also stress that this approach is well suitable for translationally invariant states where intensive variables are unambiguously representative of the global state of the system.
The equations of motion for the populations read as
\begin{equation}
\label{su3}
\dot n_{\mu} = -\gamma \ n_{\mu} \sum_{\mu'\neq\mu} n_{\mu'}.
\end{equation}
The dynamics cannot be analytically solved for a generic initial condition but only in few cases that we will now discuss.  
When the system is initially prepared with a large and equal fraction of the total population in the $A$ and $B$ sector and only a small amount of particles in the $C$ subspace, $n_{C}(0)\ll n_{A}(0)=n_{B}(0)= \mathcal{O}(1)$ the exact solution at first order in $n_{C}(0)$ reads~\cite{SuppMat}
\begin{eqnarray}
\label{exactCsmall}
n_{A,B}(t) &=& \frac{n_{A,B} (0) }{1+\gamma t  \ n_{A,B} (0)} -  \frac{n_{C}(0) \ln[1+\gamma t  \ n_{A,B} (0))]}{\left[ 1+\gamma t  \ n_{A,B} (0)\right]^{2}}, \cr
n_{C}(t) &=&  \frac{n_{C}(0)}{\left[ 1+\gamma t  \ n_{A,B} (0)\right]^{2}}.
\end{eqnarray} 
We found that the system gets empty in the long-time limit, i.e. $\lim_{t\to\infty}n_{A,B,C}=0$.
This is expected in the $A,B$ sector since the initial condition $s_{AB}(0)=0$ implies $s_{AB}(t)=0, \forall t>0$ and the system must evolve toward the origin of the Dicke cone  $s_{AB}=n_{AB}=0$.
In the $A,C$ (or equivalently $B,C$) sectors the situation is quite different since we start from a large value of the spin $s_{AC}=s_{AC}^{z}=\hbar(n_{A}-n_{C})/2$ and again we flow toward
the vacuum. This dynamics is shown in Fig.~\ref{Fig:SU3_case12} (top panel) and the numerics shows a good agreement with the perturbative prediction~\eqref{exactCsmall}.
We also note that the presence of a non-vanishing population in $C$ modifies the $1/t$ mean-field-like decay of $n_{A,B}$ and determines a $1/t^{2}$ decay for $n_{C}$.

We now consider the situation where the system is initially prepared with a large fraction of the total population in the $A$ sector and a small (and equal) fraction of particles in the $B,C$ sectors, i.e. $n_{B}(0)=n_{C}(0)\ll n_{A}(0)$. At first order in $n_{C}(0)$ we find~\cite{SuppMat}
\begin{eqnarray}
\label{su3limB}
n_{A}(t) &=& n_{A}(0) -2 n_{B}(0)\left(1-e^{-\gamma n_{A}(0) t}\right),\cr
n_{B,C}(t) &=& n_{B,C}(0) \ e^{-\gamma n_{A}(0) t}.
\end{eqnarray}
In this case we get a steady-state with a non-vanishing particle density in the $A$ sector, i.e. $\lim_{t\to\infty}n_{A}(t)=n_{A}(0)-2n_{B}(0)$, while the $B,C$ sectors get empty $\lim_{t\to\infty}n_{B,C}(t)=0$.
This determines a non-trivial dynamics in the $AB$ subspace as shown in Fig.~\ref{Fig:SU3_case12} (bottom panel), which is well captured by Eq.~\eqref{su3limB} for small values of $n_{C}$. 

We conclude this part considering the case of equally-populated spin sectors. 
This state is of particular interest since can be easily realized in experiments~\cite{Sponselee_2018} and corresponds to a product state in which we have one particle per lattice site with maximally-mixed spin degrees of freedom. We dubbed this state {\it Mott incoherent state}. 
This state has a total spin that vanishes in the thermodynamic limit as $s^{2}_{\mu\mu'}\sim1/L \ \forall \mu\neq\mu'$.
In this case Eq.~\eqref{su3} leads to $\dot n(t) = -\gamma(N-1)n^{2}(t)$ which is solved for $n(t)/n(0) = (1+t\gamma n(0) (N-1))^{-1}$.
Here, the populations decay as $1/t$ with a typical rate given by $\gamma n(0) (N-1)$.

\textit{Strongly interacting and strongly dissipative limit--}
\begin{figure}[t]
\includegraphics[width=0.7\columnwidth]{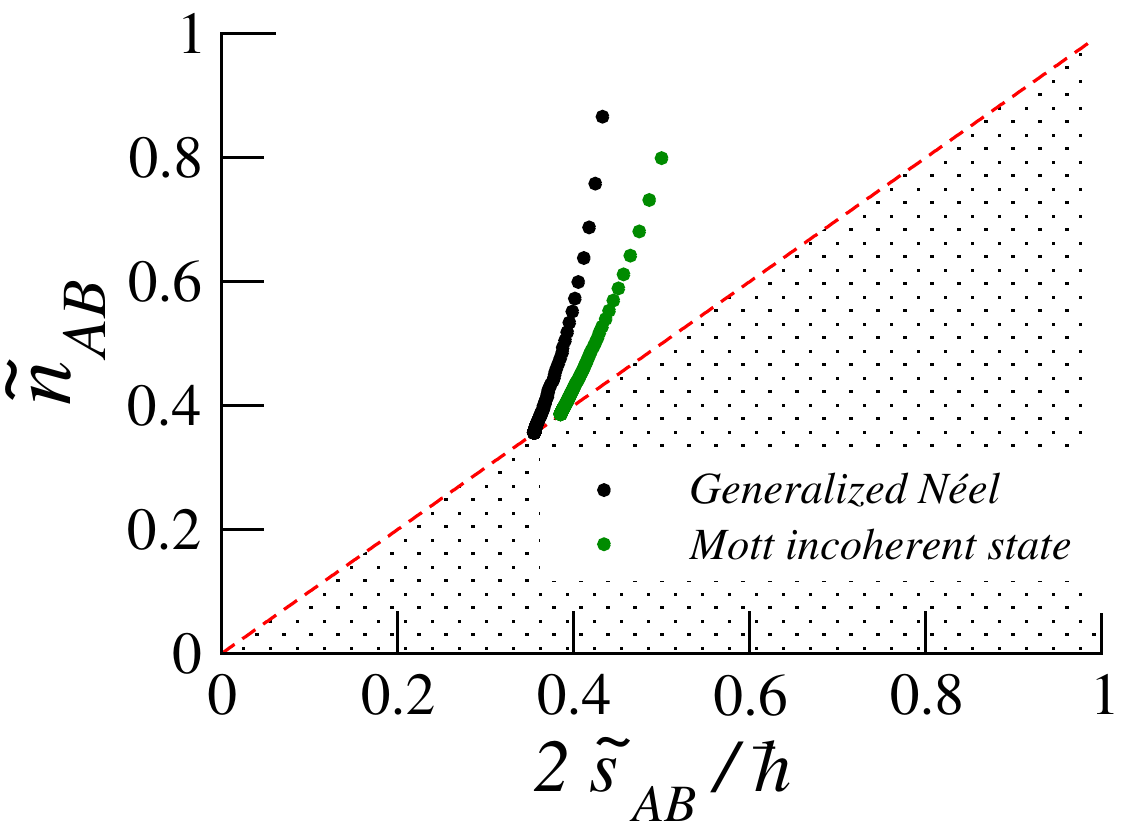}
\caption{SU($3$) dynamics in the $\tilde{s}_{AB} - \tilde{n}_{AB}$ plane. Orange circles: dynamics from the generalized N\'eel state. Green squares: dyanmics from the Mott incoherent state. The dashed line represents the Dicke cone satisfying Eq.~\eqref{dickecondition}.}
\label{Fig:grace_SUN}
\end{figure}
Let us now consider the strongly interacting and dissipative limit in which $\hbar \gamma,U \gg J$. In this limit states with at most one particle per lattice site are quasi-stationary while states with more than one excitation per lattice site are energetically disfavoured and will be quickly dissipated on a timescale proportional to $1/(\hbar \gamma)$. Consequently, the dynamics at long times will mainly take place in the hard-core fermion subspace with a new relevant timescale, namely $\Gamma_{\rm eff} \sim 1/\gamma$, which is inversely proportional to the original dissipation rate, a typical signature of the Quantum Zeno effect~\cite{Rosso_2021,Rossini_2020}. 

Following a method first proposed in Ref.~\cite{GarciaRipoll_2009}, we derive an effective Lindblad master equation that governs the dynamics in this regime~\cite{SuppMat}.
The effective Hamiltonian $\hat H' = -J \sum_{j, \mu} (\hat f_{j, \mu}^\dagger \hat f_{j+1, \mu}+ {\rm H.c.})$ corresponds to a hopping Hamiltonian of hardcore fermions annihilated by the operators $\hat f_{j, \mu}$. The effective jump operator takes into account nearest-neighbor losses $\hat L'_{j, \mu \mu'} = \sqrt{\Gamma_{\rm eff}} [\hat f_{j \mu} (\hat f_{j-1, \mu'} + \hat f_{j+1, \mu'} ) - \hat f_{j \mu'} (\hat f_{j-1, \mu} + \hat f_{j+1, \mu} ) ]$, with $\mu<\mu'$ and 
$
\Gamma_{\rm eff} = \frac{4}{1 + \left( \frac{2U}{\hbar \gamma} \right)^2 } \frac{J^2}{\hbar^2 \gamma}$.
This effective master equation is a generalization to the SU($3$) case of the one presented in Ref.~\cite{Zhu_2014}.
We want now to show that also in this regime the steady-state is a mixture of generalized Dicke state. As a smoking gun we will study whether the condition \eqref{dickecondition}
holds at long times for a generic $\mu,\mu'$ subspace. In order to verify this relation we solve numerically the master equation for open boundary conditions by means of quantum trajectories~\cite{Qutip01,Qutip02}. In particular, we consider the dynamics starting from a generalized N\'eel state of the form $\ket{\psi_{\rm g- Neel}} = \ket{A \; \; B \; \; C \dots A \; \; B \; \; C}$ and the Mott incoherent state.  
In Fig.~\ref{Fig:grace_SUN} we plot the system evolution in the $AB$ subspace in the $2\tilde{s}_{AB}/\hbar - \tilde{n}_{AB}$ plane where we defined 
\begin{equation}
\tilde{s}_{AB}=\frac{\sqrt{\langle\hat S^{2}_{AB}\rangle}}{L}, \ \ \tilde{n}_{AB}=\frac{2\sqrt{\langle\frac{\hat N_{AB}}{2} (\frac{\hat N_{AB}}{2} +1)}\rangle}{L}.
\end{equation}
Again, in the long-time limit the curves asymptotically collapse on the Dicke cone where Eq.~\eqref{dickecondition} holds. The latter statement is true for any of the subspaces; for what concerns the Mott incoherent state, given its particular structure and symmetry, we have that the dynamics is the same in each of the subspaces~\cite{SuppMat}. 

\textbf{Conclusions--}
In this paper we studied the dynamics and steady-state properties of a SU($3$)-symmetric cold-atom system in presence of two-body losses. While we explicitly considered the $N=3$ case, our results are qualitatively valid for any $N>2$, included $N=6$ for which experiments have been performed~\cite{Sponselee_2018}.
This work also paves the way to future intriguing research directions. Among them we mention the study of inhomogeneous situations where the tensor $\mathsf{G}(x)$ acquires a spatial dependence allowing the exploration of analogies with general relativity and the implementation of experimentally-friendly protocols for the certification and exploitation of generalized Dicke states. 

\textbf{Acknowledgments--} 
We warmly acknowledge enlightening discussions with C. Becker, K. Sponselee and J. De Nardis.
This work has been partially funded by LabEx PALM (ANR-10-LABX-0039-PALM).

\bibliography{SUNdiss.bib}

\begin{thebibliography}{49}%
\makeatletter
\providecommand \@ifxundefined [1]{%
 \@ifx{#1\undefined}
}%
\providecommand \@ifnum [1]{%
 \ifnum #1\expandafter \@firstoftwo
 \else \expandafter \@secondoftwo
 \fi
}%
\providecommand \@ifx [1]{%
 \ifx #1\expandafter \@firstoftwo
 \else \expandafter \@secondoftwo
 \fi
}%
\providecommand \natexlab [1]{#1}%
\providecommand \enquote  [1]{``#1''}%
\providecommand \bibnamefont  [1]{#1}%
\providecommand \bibfnamefont [1]{#1}%
\providecommand \citenamefont [1]{#1}%
\providecommand \href@noop [0]{\@secondoftwo}%
\providecommand \href [0]{\begingroup \@sanitize@url \@href}%
\providecommand \@href[1]{\@@startlink{#1}\@@href}%
\providecommand \@@href[1]{\endgroup#1\@@endlink}%
\providecommand \@sanitize@url [0]{\catcode `\\12\catcode `\$12\catcode
  `\&12\catcode `\#12\catcode `\^12\catcode `\_12\catcode `\%12\relax}%
\providecommand \@@startlink[1]{}%
\providecommand \@@endlink[0]{}%
\providecommand \url  [0]{\begingroup\@sanitize@url \@url }%
\providecommand \@url [1]{\endgroup\@href {#1}{\urlprefix }}%
\providecommand \urlprefix  [0]{URL }%
\providecommand \Eprint [0]{\href }%
\providecommand \doibase [0]{https://doi.org/}%
\providecommand \selectlanguage [0]{\@gobble}%
\providecommand \bibinfo  [0]{\@secondoftwo}%
\providecommand \bibfield  [0]{\@secondoftwo}%
\providecommand \translation [1]{[#1]}%
\providecommand \BibitemOpen [0]{}%
\providecommand \bibitemStop [0]{}%
\providecommand \bibitemNoStop [0]{.\EOS\space}%
\providecommand \EOS [0]{\spacefactor3000\relax}%
\providecommand \BibitemShut  [1]{\csname bibitem#1\endcsname}%
\let\auto@bib@innerbib\@empty
\bibitem [{\citenamefont {Langen}\ \emph {et~al.}(2015)\citenamefont {Langen},
  \citenamefont {Geiger},\ and\ \citenamefont {Schmiedmayer}}]{LangenRev_2015}%
  \BibitemOpen
  \bibfield  {author} {\bibinfo {author} {\bibfnamefont {T.}~\bibnamefont
  {Langen}}, \bibinfo {author} {\bibfnamefont {R.}~\bibnamefont {Geiger}},\
  and\ \bibinfo {author} {\bibfnamefont {J.}~\bibnamefont {Schmiedmayer}},\
  }\bibfield  {title} {\bibinfo {title} {Ultracold atoms out of equilibrium},\
  }\href {https://doi.org/10.1146/annurev-conmatphys-031214-014548} {\bibfield
  {journal} {\bibinfo  {journal} {Annual Review of Condensed Matter Physics}\
  }\textbf {\bibinfo {volume} {6}},\ \bibinfo {pages} {201} (\bibinfo {year}
  {2015})}\BibitemShut {NoStop}%
\bibitem [{\citenamefont {Gross}\ and\ \citenamefont
  {Bloch}(2017)}]{Gross2017}%
  \BibitemOpen
  \bibfield  {author} {\bibinfo {author} {\bibfnamefont {C.}~\bibnamefont
  {Gross}}\ and\ \bibinfo {author} {\bibfnamefont {I.}~\bibnamefont {Bloch}},\
  }\bibfield  {title} {\bibinfo {title} {Quantum simulations with ultracold
  atoms in optical lattices},\ }\href {https://doi.org/10.1126/science.aal3837}
  {\bibfield  {journal} {\bibinfo  {journal} {Science}\ }\textbf {\bibinfo
  {volume} {357}},\ \bibinfo {pages} {995} (\bibinfo {year} {2017})},\ \Eprint
  {https://arxiv.org/abs/https://www.science.org/doi/pdf/10.1126/science.aal3837}
  {https://www.science.org/doi/pdf/10.1126/science.aal3837} \BibitemShut
  {NoStop}%
\bibitem [{\citenamefont {Sch{\"a}fer}\ \emph {et~al.}(2020)\citenamefont
  {Sch{\"a}fer}, \citenamefont {Fukuhara}, \citenamefont {Sugawa},
  \citenamefont {Takasu},\ and\ \citenamefont {Takahashi}}]{Schafer2020}%
  \BibitemOpen
  \bibfield  {author} {\bibinfo {author} {\bibfnamefont {F.}~\bibnamefont
  {Sch{\"a}fer}}, \bibinfo {author} {\bibfnamefont {T.}~\bibnamefont
  {Fukuhara}}, \bibinfo {author} {\bibfnamefont {S.}~\bibnamefont {Sugawa}},
  \bibinfo {author} {\bibfnamefont {Y.}~\bibnamefont {Takasu}},\ and\ \bibinfo
  {author} {\bibfnamefont {Y.}~\bibnamefont {Takahashi}},\ }\bibfield  {title}
  {\bibinfo {title} {Tools for quantum simulation with ultracold atoms in
  optical lattices},\ }\href {https://doi.org/10.1038/s42254-020-0195-3}
  {\bibfield  {journal} {\bibinfo  {journal} {Nature Reviews Physics}\ }\textbf
  {\bibinfo {volume} {2}},\ \bibinfo {pages} {411} (\bibinfo {year}
  {2020})}\BibitemShut {NoStop}%
\bibitem [{\citenamefont {Zurek}(2003)}]{Zurek_2003}%
  \BibitemOpen
  \bibfield  {author} {\bibinfo {author} {\bibfnamefont {W.~H.}\ \bibnamefont
  {Zurek}},\ }\bibfield  {title} {\bibinfo {title} {Decoherence, einselection,
  and the quantum origins of the classical},\ }\href
  {https://doi.org/10.1103/RevModPhys.75.715} {\bibfield  {journal} {\bibinfo
  {journal} {Rev. Mod. Phys.}\ }\textbf {\bibinfo {volume} {75}},\ \bibinfo
  {pages} {715} (\bibinfo {year} {2003})}\BibitemShut {NoStop}%
\bibitem [{\citenamefont {Syassen}\ \emph {et~al.}(2008)\citenamefont
  {Syassen}, \citenamefont {Bauer}, \citenamefont {Lettner}, \citenamefont
  {Volz}, \citenamefont {Dietze}, \citenamefont {Garc{\'\i}a-Ripoll},
  \citenamefont {Cirac}, \citenamefont {Rempe},\ and\ \citenamefont
  {D{\"u}rr}}]{Syassen_2008}%
  \BibitemOpen
  \bibfield  {author} {\bibinfo {author} {\bibfnamefont {N.}~\bibnamefont
  {Syassen}}, \bibinfo {author} {\bibfnamefont {D.~M.}\ \bibnamefont {Bauer}},
  \bibinfo {author} {\bibfnamefont {M.}~\bibnamefont {Lettner}}, \bibinfo
  {author} {\bibfnamefont {T.}~\bibnamefont {Volz}}, \bibinfo {author}
  {\bibfnamefont {D.}~\bibnamefont {Dietze}}, \bibinfo {author} {\bibfnamefont
  {J.~J.}\ \bibnamefont {Garc{\'\i}a-Ripoll}}, \bibinfo {author} {\bibfnamefont
  {J.~I.}\ \bibnamefont {Cirac}}, \bibinfo {author} {\bibfnamefont
  {G.}~\bibnamefont {Rempe}},\ and\ \bibinfo {author} {\bibfnamefont
  {S.}~\bibnamefont {D{\"u}rr}},\ }\bibfield  {title} {\bibinfo {title} {Strong
  dissipation inhibits losses and induces correlations in cold molecular
  gases},\ }\href {https://doi.org/10.1126/science.1155309} {\bibfield
  {journal} {\bibinfo  {journal} {Science}\ }\textbf {\bibinfo {volume}
  {320}},\ \bibinfo {pages} {1329} (\bibinfo {year} {2008})}\BibitemShut
  {NoStop}%
\bibitem [{\citenamefont {Garc{\'{\i}}a-Ripoll}\ \emph
  {et~al.}(2009)\citenamefont {Garc{\'{\i}}a-Ripoll}, \citenamefont {D\"urr},
  \citenamefont {Syassen}, \citenamefont {Bauer}, \citenamefont {Lettner},
  \citenamefont {Rempe},\ and\ \citenamefont {Cirac}}]{GarciaRipoll_2009}%
  \BibitemOpen
  \bibfield  {author} {\bibinfo {author} {\bibfnamefont {J.~J.}\ \bibnamefont
  {Garc{\'{\i}}a-Ripoll}}, \bibinfo {author} {\bibfnamefont {S.}~\bibnamefont
  {D\"urr}}, \bibinfo {author} {\bibfnamefont {N.}~\bibnamefont {Syassen}},
  \bibinfo {author} {\bibfnamefont {D.~M.}\ \bibnamefont {Bauer}}, \bibinfo
  {author} {\bibfnamefont {M.}~\bibnamefont {Lettner}}, \bibinfo {author}
  {\bibfnamefont {G.}~\bibnamefont {Rempe}},\ and\ \bibinfo {author}
  {\bibfnamefont {J.~I.}\ \bibnamefont {Cirac}},\ }\bibfield  {title} {\bibinfo
  {title} {Dissipation-induced hard-core boson gas in an optical lattice},\
  }\href {https://doi.org/10.1088/1367-2630/11/1/013053} {\bibfield  {journal}
  {\bibinfo  {journal} {New J. Phys.}\ }\textbf {\bibinfo {volume} {11}},\
  \bibinfo {pages} {013053} (\bibinfo {year} {2009})}\BibitemShut {NoStop}%
\bibitem [{\citenamefont {Kantian}\ \emph {et~al.}(2009)\citenamefont
  {Kantian}, \citenamefont {Dalmonte}, \citenamefont {Diehl}, \citenamefont
  {Hofstetter}, \citenamefont {Zoller},\ and\ \citenamefont
  {Daley}}]{Kantian_2009}%
  \BibitemOpen
  \bibfield  {author} {\bibinfo {author} {\bibfnamefont {A.}~\bibnamefont
  {Kantian}}, \bibinfo {author} {\bibfnamefont {M.}~\bibnamefont {Dalmonte}},
  \bibinfo {author} {\bibfnamefont {S.}~\bibnamefont {Diehl}}, \bibinfo
  {author} {\bibfnamefont {W.}~\bibnamefont {Hofstetter}}, \bibinfo {author}
  {\bibfnamefont {P.}~\bibnamefont {Zoller}},\ and\ \bibinfo {author}
  {\bibfnamefont {A.~J.}\ \bibnamefont {Daley}},\ }\bibfield  {title} {\bibinfo
  {title} {Atomic color superfluid via three-body loss},\ }\href
  {https://doi.org/10.1103/PhysRevLett.103.240401} {\bibfield  {journal}
  {\bibinfo  {journal} {Phys. Rev. Lett.}\ }\textbf {\bibinfo {volume} {103}},\
  \bibinfo {pages} {240401} (\bibinfo {year} {2009})}\BibitemShut {NoStop}%
\bibitem [{\citenamefont {Letscher}\ \emph {et~al.}(2017)\citenamefont
  {Letscher}, \citenamefont {Thomas}, \citenamefont {Niederpr\"um},
  \citenamefont {Fleischhauer},\ and\ \citenamefont {Ott}}]{Letscher_2017}%
  \BibitemOpen
  \bibfield  {author} {\bibinfo {author} {\bibfnamefont {F.}~\bibnamefont
  {Letscher}}, \bibinfo {author} {\bibfnamefont {O.}~\bibnamefont {Thomas}},
  \bibinfo {author} {\bibfnamefont {T.}~\bibnamefont {Niederpr\"um}}, \bibinfo
  {author} {\bibfnamefont {M.}~\bibnamefont {Fleischhauer}},\ and\ \bibinfo
  {author} {\bibfnamefont {H.}~\bibnamefont {Ott}},\ }\bibfield  {title}
  {\bibinfo {title} {Bistability versus metastability in driven dissipative
  rydberg gases},\ }\href {https://doi.org/10.1103/PhysRevX.7.021020}
  {\bibfield  {journal} {\bibinfo  {journal} {Phys. Rev. X}\ }\textbf {\bibinfo
  {volume} {7}},\ \bibinfo {pages} {021020} (\bibinfo {year}
  {2017})}\BibitemShut {NoStop}%
\bibitem [{\citenamefont {Diehl}\ \emph {et~al.}(2008)\citenamefont {Diehl},
  \citenamefont {Micheli}, \citenamefont {Kantian}, \citenamefont {Kraus},
  \citenamefont {B\"uchler},\ and\ \citenamefont {Zoller}}]{Diehl_2008}%
  \BibitemOpen
  \bibfield  {author} {\bibinfo {author} {\bibfnamefont {S.}~\bibnamefont
  {Diehl}}, \bibinfo {author} {\bibfnamefont {A.}~\bibnamefont {Micheli}},
  \bibinfo {author} {\bibfnamefont {A.}~\bibnamefont {Kantian}}, \bibinfo
  {author} {\bibfnamefont {B.}~\bibnamefont {Kraus}}, \bibinfo {author}
  {\bibfnamefont {H.-P.}\ \bibnamefont {B\"uchler}},\ and\ \bibinfo {author}
  {\bibfnamefont {P.}~\bibnamefont {Zoller}},\ }\bibfield  {title} {\bibinfo
  {title} {Quantum states and phases in driven open quantum systems with cold
  atoms},\ }\href {https://doi.org/10.1038/nphys1073} {\bibfield  {journal}
  {\bibinfo  {journal} {Nat. Phys.}\ }\textbf {\bibinfo {volume} {4}},\
  \bibinfo {pages} {878} (\bibinfo {year} {2008})}\BibitemShut {NoStop}%
\bibitem [{\citenamefont {Lee}\ \emph {et~al.}(2011)\citenamefont {Lee},
  \citenamefont {Haffner},\ and\ \citenamefont {Cross}}]{Lee_2011}%
  \BibitemOpen
  \bibfield  {author} {\bibinfo {author} {\bibfnamefont {T.~E.}\ \bibnamefont
  {Lee}}, \bibinfo {author} {\bibfnamefont {H.}~\bibnamefont {Haffner}},\ and\
  \bibinfo {author} {\bibfnamefont {M.~C.}\ \bibnamefont {Cross}},\ }\bibfield
  {title} {\bibinfo {title} {Antiferromagnetic phase transition in a
  nonequilibrium lattice of rydberg atoms},\ }\href@noop {} {\bibfield
  {journal} {\bibinfo  {journal} {Physical Review A}\ }\textbf {\bibinfo
  {volume} {84}},\ \bibinfo {pages} {031402} (\bibinfo {year}
  {2011})}\BibitemShut {NoStop}%
\bibitem [{\citenamefont {Jin}\ \emph {et~al.}(2016)\citenamefont {Jin},
  \citenamefont {Biella}, \citenamefont {Viyuela}, \citenamefont {Mazza},
  \citenamefont {Keeling}, \citenamefont {Fazio},\ and\ \citenamefont
  {Rossini}}]{Jin_2016}%
  \BibitemOpen
  \bibfield  {author} {\bibinfo {author} {\bibfnamefont {J.}~\bibnamefont
  {Jin}}, \bibinfo {author} {\bibfnamefont {A.}~\bibnamefont {Biella}},
  \bibinfo {author} {\bibfnamefont {O.}~\bibnamefont {Viyuela}}, \bibinfo
  {author} {\bibfnamefont {L.}~\bibnamefont {Mazza}}, \bibinfo {author}
  {\bibfnamefont {J.}~\bibnamefont {Keeling}}, \bibinfo {author} {\bibfnamefont
  {R.}~\bibnamefont {Fazio}},\ and\ \bibinfo {author} {\bibfnamefont
  {D.}~\bibnamefont {Rossini}},\ }\bibfield  {title} {\bibinfo {title} {Cluster
  mean-field approach to the steady-state phase diagram of dissipative spin
  systems},\ }\href {https://doi.org/10.1103/PhysRevX.6.031011} {\bibfield
  {journal} {\bibinfo  {journal} {Phys. Rev. X}\ }\textbf {\bibinfo {volume}
  {6}},\ \bibinfo {pages} {031011} (\bibinfo {year} {2016})}\BibitemShut
  {NoStop}%
\bibitem [{\citenamefont {Morsch}\ and\ \citenamefont
  {Lesanovsky}(2018)}]{Morsch_2018}%
  \BibitemOpen
  \bibfield  {author} {\bibinfo {author} {\bibfnamefont {O.}~\bibnamefont
  {Morsch}}\ and\ \bibinfo {author} {\bibfnamefont {I.}~\bibnamefont
  {Lesanovsky}},\ }\bibfield  {title} {\bibinfo {title} {Dissipative many-body
  physics of cold rydberg atoms},\ }\href
  {https://doi.org/10.1393/ncr/i2018-10149-7} {\bibfield  {journal} {\bibinfo
  {journal} {La Rivista del Nuovo Cimento}\ }\textbf {\bibinfo {volume} {41}},\
  \bibinfo {pages} {383} (\bibinfo {year} {2018})}\BibitemShut {NoStop}%
\bibitem [{\citenamefont {Verstraete}\ \emph {et~al.}(2009)\citenamefont
  {Verstraete}, \citenamefont {Wolf},\ and\ \citenamefont
  {Cirac}}]{Verstraete_2009}%
  \BibitemOpen
  \bibfield  {author} {\bibinfo {author} {\bibfnamefont {F.}~\bibnamefont
  {Verstraete}}, \bibinfo {author} {\bibfnamefont {M.~M.}\ \bibnamefont
  {Wolf}},\ and\ \bibinfo {author} {\bibfnamefont {J.~I.}\ \bibnamefont
  {Cirac}},\ }\bibfield  {title} {\bibinfo {title} {Quantum computation,
  quantum state engineering, and quantum phase transitions driven by
  dissipation},\ }\href {https://doi.org/10.1038/nphys1342} {\bibfield
  {journal} {\bibinfo  {journal} {Nat. Phys.}\ }\textbf {\bibinfo {volume}
  {5}},\ \bibinfo {pages} {633} (\bibinfo {year} {2009})}\BibitemShut {NoStop}%
\bibitem [{\citenamefont {Sponselee}\ \emph {et~al.}(2019)\citenamefont
  {Sponselee}, \citenamefont {Freystatzky}, \citenamefont {Abeln},
  \citenamefont {Diem}, \citenamefont {Hundt}, \citenamefont {Kochanke},
  \citenamefont {Ponath}, \citenamefont {Santra}, \citenamefont {Mathey},
  \citenamefont {Sengstock},\ and\ \citenamefont {Becker}}]{Sponselee_2018}%
  \BibitemOpen
  \bibfield  {author} {\bibinfo {author} {\bibfnamefont {K.}~\bibnamefont
  {Sponselee}}, \bibinfo {author} {\bibfnamefont {L.}~\bibnamefont
  {Freystatzky}}, \bibinfo {author} {\bibfnamefont {B.}~\bibnamefont {Abeln}},
  \bibinfo {author} {\bibfnamefont {M.}~\bibnamefont {Diem}}, \bibinfo {author}
  {\bibfnamefont {B.}~\bibnamefont {Hundt}}, \bibinfo {author} {\bibfnamefont
  {A.}~\bibnamefont {Kochanke}}, \bibinfo {author} {\bibfnamefont
  {T.}~\bibnamefont {Ponath}}, \bibinfo {author} {\bibfnamefont
  {B.}~\bibnamefont {Santra}}, \bibinfo {author} {\bibfnamefont
  {L.}~\bibnamefont {Mathey}}, \bibinfo {author} {\bibfnamefont
  {K.}~\bibnamefont {Sengstock}},\ and\ \bibinfo {author} {\bibfnamefont
  {C.}~\bibnamefont {Becker}},\ }\bibfield  {title} {\bibinfo {title} {Dynamics
  of ultracold quantum gases in the dissipative fermi-hubbard model},\ }\href
  {https://doi.org/10.1088/2058-9565/aadccd} {\bibfield  {journal} {\bibinfo
  {journal} {Quantum Sci. Technol.}\ }\textbf {\bibinfo {volume} {4}},\
  \bibinfo {pages} {014002} (\bibinfo {year} {2019})}\BibitemShut {NoStop}%
\bibitem [{\citenamefont {Rosso}\ \emph {et~al.}(2021)\citenamefont {Rosso},
  \citenamefont {Rossini}, \citenamefont {Biella},\ and\ \citenamefont
  {Mazza}}]{Rosso_2021}%
  \BibitemOpen
  \bibfield  {author} {\bibinfo {author} {\bibfnamefont {L.}~\bibnamefont
  {Rosso}}, \bibinfo {author} {\bibfnamefont {D.}~\bibnamefont {Rossini}},
  \bibinfo {author} {\bibfnamefont {A.}~\bibnamefont {Biella}},\ and\ \bibinfo
  {author} {\bibfnamefont {L.}~\bibnamefont {Mazza}},\ }\bibfield  {title}
  {\bibinfo {title} {One-dimensional spin-1/2 fermionic gases with two-body
  losses: Weak dissipation and spin conservation},\ }\href
  {https://doi.org/10.1103/PhysRevA.104.053305} {\bibfield  {journal} {\bibinfo
   {journal} {Phys. Rev. A}\ }\textbf {\bibinfo {volume} {104}},\ \bibinfo
  {pages} {053305} (\bibinfo {year} {2021})}\BibitemShut {NoStop}%
\bibitem [{\citenamefont {Kordas}\ \emph {et~al.}(2015)\citenamefont {Kordas},
  \citenamefont {Witthaut}, \citenamefont {Buonsante}, \citenamefont {Vezzani},
  \citenamefont {Burioni}, \citenamefont {Karanikas},\ and\ \citenamefont
  {Wimberger}}]{Kordas_2015}%
  \BibitemOpen
  \bibfield  {author} {\bibinfo {author} {\bibfnamefont {G.}~\bibnamefont
  {Kordas}}, \bibinfo {author} {\bibfnamefont {D.}~\bibnamefont {Witthaut}},
  \bibinfo {author} {\bibfnamefont {P.}~\bibnamefont {Buonsante}}, \bibinfo
  {author} {\bibfnamefont {A.}~\bibnamefont {Vezzani}}, \bibinfo {author}
  {\bibfnamefont {R.}~\bibnamefont {Burioni}}, \bibinfo {author} {\bibfnamefont
  {A.~I.}\ \bibnamefont {Karanikas}},\ and\ \bibinfo {author} {\bibfnamefont
  {S.}~\bibnamefont {Wimberger}},\ }\bibfield  {title} {\bibinfo {title} {The
  dissipative bose-hubbard model},\ }\href
  {https://doi.org/10.1140/epjst/e2015-02528-2} {\bibfield  {journal} {\bibinfo
   {journal} {The European Physical Journal Special Topics}\ }\textbf {\bibinfo
  {volume} {224}},\ \bibinfo {pages} {2127} (\bibinfo {year}
  {2015})}\BibitemShut {NoStop}%
\bibitem [{\citenamefont {Johnson}\ \emph {et~al.}(2017)\citenamefont
  {Johnson}, \citenamefont {Szigeti}, \citenamefont {Schemmer},\ and\
  \citenamefont {Bouchoule}}]{Johnson_2017}%
  \BibitemOpen
  \bibfield  {author} {\bibinfo {author} {\bibfnamefont {A.}~\bibnamefont
  {Johnson}}, \bibinfo {author} {\bibfnamefont {S.~S.}\ \bibnamefont
  {Szigeti}}, \bibinfo {author} {\bibfnamefont {M.}~\bibnamefont {Schemmer}},\
  and\ \bibinfo {author} {\bibfnamefont {I.}~\bibnamefont {Bouchoule}},\
  }\bibfield  {title} {\bibinfo {title} {Long-lived nonthermal states realized
  by atom losses in one-dimensional quasicondensates},\ }\href
  {https://doi.org/10.1103/PhysRevA.96.013623} {\bibfield  {journal} {\bibinfo
  {journal} {Phys. Rev. A}\ }\textbf {\bibinfo {volume} {96}},\ \bibinfo
  {pages} {013623} (\bibinfo {year} {2017})}\BibitemShut {NoStop}%
\bibitem [{\citenamefont {Schemmer}\ and\ \citenamefont
  {Bouchoule}(2018)}]{Schemmer_2018}%
  \BibitemOpen
  \bibfield  {author} {\bibinfo {author} {\bibfnamefont {M.}~\bibnamefont
  {Schemmer}}\ and\ \bibinfo {author} {\bibfnamefont {I.}~\bibnamefont
  {Bouchoule}},\ }\bibfield  {title} {\bibinfo {title} {Cooling a bose gas by
  three-body losses},\ }\href {https://doi.org/10.1103/PhysRevLett.121.200401}
  {\bibfield  {journal} {\bibinfo  {journal} {Phys. Rev. Lett.}\ }\textbf
  {\bibinfo {volume} {121}},\ \bibinfo {pages} {200401} (\bibinfo {year}
  {2018})}\BibitemShut {NoStop}%
\bibitem [{\citenamefont {Bouchoule}\ \emph {et~al.}(2020)\citenamefont
  {Bouchoule}, \citenamefont {Doyon},\ and\ \citenamefont
  {Dubail}}]{Bouchoule_2020b}%
  \BibitemOpen
  \bibfield  {author} {\bibinfo {author} {\bibfnamefont {I.}~\bibnamefont
  {Bouchoule}}, \bibinfo {author} {\bibfnamefont {B.}~\bibnamefont {Doyon}},\
  and\ \bibinfo {author} {\bibfnamefont {J.}~\bibnamefont {Dubail}},\
  }\bibfield  {title} {\bibinfo {title} {{The effect of atom losses on the
  distribution of rapidities in the one-dimensional Bose gas}},\ }\href
  {https://doi.org/10.21468/SciPostPhys.9.4.044} {\bibfield  {journal}
  {\bibinfo  {journal} {SciPost Phys.}\ }\textbf {\bibinfo {volume} {9}},\
  \bibinfo {pages} {44} (\bibinfo {year} {2020})}\BibitemShut {NoStop}%
\bibitem [{\citenamefont {Bouchoule}\ and\ \citenamefont
  {Schemmer}(2020)}]{Bouchoule_2020SCP}%
  \BibitemOpen
  \bibfield  {author} {\bibinfo {author} {\bibfnamefont {I.}~\bibnamefont
  {Bouchoule}}\ and\ \bibinfo {author} {\bibfnamefont {M.}~\bibnamefont
  {Schemmer}},\ }\bibfield  {title} {\bibinfo {title} {{Asymptotic temperature
  of a lossy condensate}},\ }\href
  {https://doi.org/10.21468/SciPostPhys.8.4.060} {\bibfield  {journal}
  {\bibinfo  {journal} {SciPost Phys.}\ }\textbf {\bibinfo {volume} {8}},\
  \bibinfo {pages} {60} (\bibinfo {year} {2020})}\BibitemShut {NoStop}%
\bibitem [{\citenamefont {Ashida}\ \emph {et~al.}(2020)\citenamefont {Ashida},
  \citenamefont {Gong},\ and\ \citenamefont {Ueda}}]{Ashida_2020}%
  \BibitemOpen
  \bibfield  {author} {\bibinfo {author} {\bibfnamefont {Y.}~\bibnamefont
  {Ashida}}, \bibinfo {author} {\bibfnamefont {Z.}~\bibnamefont {Gong}},\ and\
  \bibinfo {author} {\bibfnamefont {M.}~\bibnamefont {Ueda}},\ }\bibfield
  {title} {\bibinfo {title} {Non-hermitian physics},\ }\href
  {https://doi.org/10.1080/00018732.2021.1876991} {\bibfield  {journal}
  {\bibinfo  {journal} {Advances in Physics}\ }\textbf {\bibinfo {volume}
  {69}},\ \bibinfo {pages} {249} (\bibinfo {year} {2020})},\ \Eprint
  {https://arxiv.org/abs/https://doi.org/10.1080/00018732.2021.1876991}
  {https://doi.org/10.1080/00018732.2021.1876991} \BibitemShut {NoStop}%
\bibitem [{\citenamefont {Bouchoule}\ and\ \citenamefont
  {Dubail}(2021)}]{Bouchoule_2021}%
  \BibitemOpen
  \bibfield  {author} {\bibinfo {author} {\bibfnamefont {I.}~\bibnamefont
  {Bouchoule}}\ and\ \bibinfo {author} {\bibfnamefont {J.}~\bibnamefont
  {Dubail}},\ }\bibfield  {title} {\bibinfo {title} {Breakdown of tan's
  relation in lossy one-dimensional bose gases},\ }\href
  {https://doi.org/10.1103/PhysRevLett.126.160603} {\bibfield  {journal}
  {\bibinfo  {journal} {Phys. Rev. Lett.}\ }\textbf {\bibinfo {volume} {126}},\
  \bibinfo {pages} {160603} (\bibinfo {year} {2021})}\BibitemShut {NoStop}%
\bibitem [{\citenamefont {Bouchoule}\ \emph {et~al.}(2021)\citenamefont
  {Bouchoule}, \citenamefont {Dubois},\ and\ \citenamefont
  {Barbier}}]{Bouchoule_2021PRA}%
  \BibitemOpen
  \bibfield  {author} {\bibinfo {author} {\bibfnamefont {I.}~\bibnamefont
  {Bouchoule}}, \bibinfo {author} {\bibfnamefont {L.}~\bibnamefont {Dubois}},\
  and\ \bibinfo {author} {\bibfnamefont {L.-P.}\ \bibnamefont {Barbier}},\
  }\bibfield  {title} {\bibinfo {title} {Losses in interacting quantum gases:
  Ultraviolet divergence and its regularization},\ }\href
  {https://doi.org/10.1103/PhysRevA.104.L031304} {\bibfield  {journal}
  {\bibinfo  {journal} {Phys. Rev. A}\ }\textbf {\bibinfo {volume} {104}},\
  \bibinfo {pages} {L031304} (\bibinfo {year} {2021})}\BibitemShut {NoStop}%
\bibitem [{\citenamefont {Nakagawa}\ \emph {et~al.}(2020)\citenamefont
  {Nakagawa}, \citenamefont {Tsuji}, \citenamefont {Kawakami},\ and\
  \citenamefont {Ueda}}]{Nakagawa_2020}%
  \BibitemOpen
  \bibfield  {author} {\bibinfo {author} {\bibfnamefont {M.}~\bibnamefont
  {Nakagawa}}, \bibinfo {author} {\bibfnamefont {N.}~\bibnamefont {Tsuji}},
  \bibinfo {author} {\bibfnamefont {N.}~\bibnamefont {Kawakami}},\ and\
  \bibinfo {author} {\bibfnamefont {M.}~\bibnamefont {Ueda}},\ }\bibfield
  {title} {\bibinfo {title} {Dynamical sign reversal of magnetic correlations
  in dissipative hubbard models},\ }\href
  {https://doi.org/10.1103/PhysRevLett.124.147203} {\bibfield  {journal}
  {\bibinfo  {journal} {Phys. Rev. Lett.}\ }\textbf {\bibinfo {volume} {124}},\
  \bibinfo {pages} {147203} (\bibinfo {year} {2020})}\BibitemShut {NoStop}%
\bibitem [{\citenamefont {Nakagawa}\ \emph {et~al.}(2021)\citenamefont
  {Nakagawa}, \citenamefont {Kawakami},\ and\ \citenamefont
  {Ueda}}]{Nakagawa_2021}%
  \BibitemOpen
  \bibfield  {author} {\bibinfo {author} {\bibfnamefont {M.}~\bibnamefont
  {Nakagawa}}, \bibinfo {author} {\bibfnamefont {N.}~\bibnamefont {Kawakami}},\
  and\ \bibinfo {author} {\bibfnamefont {M.}~\bibnamefont {Ueda}},\ }\bibfield
  {title} {\bibinfo {title} {Exact liouvillian spectrum of a one-dimensional
  dissipative hubbard model},\ }\href
  {https://doi.org/10.1103/PhysRevLett.126.110404} {\bibfield  {journal}
  {\bibinfo  {journal} {Phys. Rev. Lett.}\ }\textbf {\bibinfo {volume} {126}},\
  \bibinfo {pages} {110404} (\bibinfo {year} {2021})}\BibitemShut {NoStop}%
\bibitem [{\citenamefont {D\"urr}\ \emph {et~al.}(2009)\citenamefont {D\"urr},
  \citenamefont {Garc\'{\i}a-Ripoll}, \citenamefont {Syassen}, \citenamefont
  {Bauer}, \citenamefont {Lettner}, \citenamefont {Cirac},\ and\ \citenamefont
  {Rempe}}]{Durr_2009}%
  \BibitemOpen
  \bibfield  {author} {\bibinfo {author} {\bibfnamefont {S.}~\bibnamefont
  {D\"urr}}, \bibinfo {author} {\bibfnamefont {J.~J.}\ \bibnamefont
  {Garc\'{\i}a-Ripoll}}, \bibinfo {author} {\bibfnamefont {N.}~\bibnamefont
  {Syassen}}, \bibinfo {author} {\bibfnamefont {D.~M.}\ \bibnamefont {Bauer}},
  \bibinfo {author} {\bibfnamefont {M.}~\bibnamefont {Lettner}}, \bibinfo
  {author} {\bibfnamefont {J.~I.}\ \bibnamefont {Cirac}},\ and\ \bibinfo
  {author} {\bibfnamefont {G.}~\bibnamefont {Rempe}},\ }\bibfield  {title}
  {\bibinfo {title} {Lieb-liniger model of a dissipation-induced
  tonks-girardeau gas},\ }\href {https://doi.org/10.1103/PhysRevA.79.023614}
  {\bibfield  {journal} {\bibinfo  {journal} {Phys. Rev. A}\ }\textbf {\bibinfo
  {volume} {79}},\ \bibinfo {pages} {023614} (\bibinfo {year}
  {2009})}\BibitemShut {NoStop}%
\bibitem [{\citenamefont {Tomita}\ \emph {et~al.}(2017)\citenamefont {Tomita},
  \citenamefont {Nakajima}, \citenamefont {Danshita}, \citenamefont {Takasu},\
  and\ \citenamefont {Takahashi}}]{Tomita_2017}%
  \BibitemOpen
  \bibfield  {author} {\bibinfo {author} {\bibfnamefont {T.}~\bibnamefont
  {Tomita}}, \bibinfo {author} {\bibfnamefont {S.}~\bibnamefont {Nakajima}},
  \bibinfo {author} {\bibfnamefont {I.}~\bibnamefont {Danshita}}, \bibinfo
  {author} {\bibfnamefont {Y.}~\bibnamefont {Takasu}},\ and\ \bibinfo {author}
  {\bibfnamefont {Y.}~\bibnamefont {Takahashi}},\ }\bibfield  {title} {\bibinfo
  {title} {Observation of the mott insulator to superfluid crossover of a
  driven-dissipative bose-hubbard system},\ }\href
  {https://doi.org/10.1126/sciadv.1701513} {\bibfield  {journal} {\bibinfo
  {journal} {Science advances}\ }\textbf {\bibinfo {volume} {3}},\ \bibinfo
  {pages} {e1701513} (\bibinfo {year} {2017})}\BibitemShut {NoStop}%
\bibitem [{\citenamefont {Tomita}\ \emph {et~al.}(2019)\citenamefont {Tomita},
  \citenamefont {Nakajima}, \citenamefont {Takasu},\ and\ \citenamefont
  {Takahashi}}]{Tomita_2019}%
  \BibitemOpen
  \bibfield  {author} {\bibinfo {author} {\bibfnamefont {T.}~\bibnamefont
  {Tomita}}, \bibinfo {author} {\bibfnamefont {S.}~\bibnamefont {Nakajima}},
  \bibinfo {author} {\bibfnamefont {Y.}~\bibnamefont {Takasu}},\ and\ \bibinfo
  {author} {\bibfnamefont {Y.}~\bibnamefont {Takahashi}},\ }\bibfield  {title}
  {\bibinfo {title} {Dissipative bose-hubbard system with intrinsic two-body
  loss},\ }\href {https://doi.org/10.1103/PhysRevA.99.031601} {\bibfield
  {journal} {\bibinfo  {journal} {Phys. Rev. A}\ }\textbf {\bibinfo {volume}
  {99}},\ \bibinfo {pages} {031601} (\bibinfo {year} {2019})}\BibitemShut
  {NoStop}%
\bibitem [{\citenamefont {Rossini}\ \emph {et~al.}(2021)\citenamefont
  {Rossini}, \citenamefont {Ghermaoui}, \citenamefont {Aguilera}, \citenamefont
  {Vatr\'e}, \citenamefont {Bouganne}, \citenamefont {Beugnon}, \citenamefont
  {Gerbier},\ and\ \citenamefont {Mazza}}]{Rossini_2020}%
  \BibitemOpen
  \bibfield  {author} {\bibinfo {author} {\bibfnamefont {D.}~\bibnamefont
  {Rossini}}, \bibinfo {author} {\bibfnamefont {A.}~\bibnamefont {Ghermaoui}},
  \bibinfo {author} {\bibfnamefont {M.~B.}\ \bibnamefont {Aguilera}}, \bibinfo
  {author} {\bibfnamefont {R.}~\bibnamefont {Vatr\'e}}, \bibinfo {author}
  {\bibfnamefont {R.}~\bibnamefont {Bouganne}}, \bibinfo {author}
  {\bibfnamefont {J.}~\bibnamefont {Beugnon}}, \bibinfo {author} {\bibfnamefont
  {F.}~\bibnamefont {Gerbier}},\ and\ \bibinfo {author} {\bibfnamefont
  {L.}~\bibnamefont {Mazza}},\ }\bibfield  {title} {\bibinfo {title} {Strong
  correlations in lossy one-dimensional quantum gases: From the quantum zeno
  effect to the generalized gibbs ensemble},\ }\href
  {https://doi.org/10.1103/PhysRevA.103.L060201} {\bibfield  {journal}
  {\bibinfo  {journal} {Phys. Rev. A}\ }\textbf {\bibinfo {volume} {103}},\
  \bibinfo {pages} {L060201} (\bibinfo {year} {2021})}\BibitemShut {NoStop}%
\bibitem [{\citenamefont {Rosso}\ \emph {et~al.}(2022)\citenamefont {Rosso},
  \citenamefont {Biella},\ and\ \citenamefont {Mazza}}]{Rosso_2021bis}%
  \BibitemOpen
  \bibfield  {author} {\bibinfo {author} {\bibfnamefont {L.}~\bibnamefont
  {Rosso}}, \bibinfo {author} {\bibfnamefont {A.}~\bibnamefont {Biella}},\ and\
  \bibinfo {author} {\bibfnamefont {L.}~\bibnamefont {Mazza}},\ }\bibfield
  {title} {\bibinfo {title} {{The one-dimensional Bose gas with strong two-body
  losses: the effect of the harmonic confinement}},\ }\href
  {https://doi.org/10.21468/SciPostPhys.12.1.044} {\bibfield  {journal}
  {\bibinfo  {journal} {SciPost Phys.}\ }\textbf {\bibinfo {volume} {12}},\
  \bibinfo {pages} {44} (\bibinfo {year} {2022})}\BibitemShut {NoStop}%
\bibitem [{\citenamefont {Scarlatella}\ \emph {et~al.}(2021)\citenamefont
  {Scarlatella}, \citenamefont {Clerk}, \citenamefont {Fazio},\ and\
  \citenamefont {Schir\'o}}]{Scarlatella_2021}%
  \BibitemOpen
  \bibfield  {author} {\bibinfo {author} {\bibfnamefont {O.}~\bibnamefont
  {Scarlatella}}, \bibinfo {author} {\bibfnamefont {A.~A.}\ \bibnamefont
  {Clerk}}, \bibinfo {author} {\bibfnamefont {R.}~\bibnamefont {Fazio}},\ and\
  \bibinfo {author} {\bibfnamefont {M.}~\bibnamefont {Schir\'o}},\ }\bibfield
  {title} {\bibinfo {title} {Dynamical mean-field theory for markovian open
  quantum many-body systems},\ }\href
  {https://doi.org/10.1103/PhysRevX.11.031018} {\bibfield  {journal} {\bibinfo
  {journal} {Phys. Rev. X}\ }\textbf {\bibinfo {volume} {11}},\ \bibinfo
  {pages} {031018} (\bibinfo {year} {2021})}\BibitemShut {NoStop}%
\bibitem [{\citenamefont {Seclì}\ \emph {et~al.}(2022)\citenamefont {Seclì},
  \citenamefont {Capone},\ and\ \citenamefont {Schirò}}]{Secli_2022}%
  \BibitemOpen
  \bibfield  {author} {\bibinfo {author} {\bibfnamefont {M.}~\bibnamefont
  {Seclì}}, \bibinfo {author} {\bibfnamefont {M.}~\bibnamefont {Capone}},\
  and\ \bibinfo {author} {\bibfnamefont {M.}~\bibnamefont {Schirò}},\
  }\href@noop {} {\bibinfo {title} {Steady-state quantum zeno effect of
  driven-dissipative bosons with dynamical mean-field theory}} (\bibinfo {year}
  {2022}),\ \Eprint {https://arxiv.org/abs/2201.03191} {arXiv:2201.03191
  [quant-ph]} \BibitemShut {NoStop}%
\bibitem [{\citenamefont {Zhu}\ \emph {et~al.}(2014)\citenamefont {Zhu},
  \citenamefont {Gadway}, \citenamefont {Foss-Feig}, \citenamefont
  {Schachenmayer}, \citenamefont {Wall}, \citenamefont {Hazzard}, \citenamefont
  {Yan}, \citenamefont {Moses}, \citenamefont {Covey}, \citenamefont {Jin},
  \citenamefont {Ye}, \citenamefont {Holland},\ and\ \citenamefont
  {Rey}}]{Zhu_2014}%
  \BibitemOpen
  \bibfield  {author} {\bibinfo {author} {\bibfnamefont {B.}~\bibnamefont
  {Zhu}}, \bibinfo {author} {\bibfnamefont {B.}~\bibnamefont {Gadway}},
  \bibinfo {author} {\bibfnamefont {M.}~\bibnamefont {Foss-Feig}}, \bibinfo
  {author} {\bibfnamefont {J.}~\bibnamefont {Schachenmayer}}, \bibinfo {author}
  {\bibfnamefont {M.~L.}\ \bibnamefont {Wall}}, \bibinfo {author}
  {\bibfnamefont {K.~R.~A.}\ \bibnamefont {Hazzard}}, \bibinfo {author}
  {\bibfnamefont {B.}~\bibnamefont {Yan}}, \bibinfo {author} {\bibfnamefont
  {S.~A.}\ \bibnamefont {Moses}}, \bibinfo {author} {\bibfnamefont {J.~P.}\
  \bibnamefont {Covey}}, \bibinfo {author} {\bibfnamefont {D.~S.}\ \bibnamefont
  {Jin}}, \bibinfo {author} {\bibfnamefont {J.}~\bibnamefont {Ye}}, \bibinfo
  {author} {\bibfnamefont {M.}~\bibnamefont {Holland}},\ and\ \bibinfo {author}
  {\bibfnamefont {A.~M.}\ \bibnamefont {Rey}},\ }\bibfield  {title} {\bibinfo
  {title} {Suppressing the loss of ultracold molecules via the continuous
  quantum zeno effect},\ }\href
  {https://doi.org/10.1103/PhysRevLett.112.070404} {\bibfield  {journal}
  {\bibinfo  {journal} {Phys. Rev. Lett.}\ }\textbf {\bibinfo {volume} {112}},\
  \bibinfo {pages} {070404} (\bibinfo {year} {2014})}\BibitemShut {NoStop}%
\bibitem [{\citenamefont {Yan}\ \emph {et~al.}()\citenamefont {Yan},
  \citenamefont {Moses}, \citenamefont {Gadway}, \citenamefont {Covey},
  \citenamefont {Hazzard}, \citenamefont {Rey}, \citenamefont {Jin},\ and\
  \citenamefont {Ye}}]{Yan_2013}%
  \BibitemOpen
  \bibfield  {author} {\bibinfo {author} {\bibfnamefont {B.}~\bibnamefont
  {Yan}}, \bibinfo {author} {\bibfnamefont {S.~A.}\ \bibnamefont {Moses}},
  \bibinfo {author} {\bibfnamefont {B.}~\bibnamefont {Gadway}}, \bibinfo
  {author} {\bibfnamefont {J.}~\bibnamefont {Covey}}, \bibinfo {author}
  {\bibfnamefont {K.~R.~A.}\ \bibnamefont {Hazzard}}, \bibinfo {author}
  {\bibfnamefont {A.~M.}\ \bibnamefont {Rey}}, \bibinfo {author} {\bibfnamefont
  {D.~S.}\ \bibnamefont {Jin}},\ and\ \bibinfo {author} {\bibfnamefont
  {J.}~\bibnamefont {Ye}},\ }\bibfield  {title} {\bibinfo {title} {Observation
  of dipolar spin-exchange interactions with lattice-confined polar
  molecules},\ }\href@noop {} {\ }\BibitemShut {NoStop}%
\bibitem [{\citenamefont {Foss-Feig}\ \emph {et~al.}(2012)\citenamefont
  {Foss-Feig}, \citenamefont {Daley}, \citenamefont {Thompson},\ and\
  \citenamefont {Rey}}]{FossFeig_2012}%
  \BibitemOpen
  \bibfield  {author} {\bibinfo {author} {\bibfnamefont {M.}~\bibnamefont
  {Foss-Feig}}, \bibinfo {author} {\bibfnamefont {A.~J.}\ \bibnamefont
  {Daley}}, \bibinfo {author} {\bibfnamefont {J.~K.}\ \bibnamefont
  {Thompson}},\ and\ \bibinfo {author} {\bibfnamefont {A.~M.}\ \bibnamefont
  {Rey}},\ }\bibfield  {title} {\bibinfo {title} {Steady-state many-body
  entanglement of hot reactive fermions},\ }\href
  {https://doi.org/10.1103/PhysRevLett.109.230501} {\bibfield  {journal}
  {\bibinfo  {journal} {Phys. Rev. Lett.}\ }\textbf {\bibinfo {volume} {109}},\
  \bibinfo {pages} {230501} (\bibinfo {year} {2012})}\BibitemShut {NoStop}%
\bibitem [{\citenamefont {Yamamoto}\ \emph {et~al.}(2019)\citenamefont
  {Yamamoto}, \citenamefont {Nakagawa}, \citenamefont {Adachi}, \citenamefont
  {Takasan}, \citenamefont {Ueda},\ and\ \citenamefont
  {Kawakami}}]{Kazuki_2019}%
  \BibitemOpen
  \bibfield  {author} {\bibinfo {author} {\bibfnamefont {K.}~\bibnamefont
  {Yamamoto}}, \bibinfo {author} {\bibfnamefont {M.}~\bibnamefont {Nakagawa}},
  \bibinfo {author} {\bibfnamefont {K.}~\bibnamefont {Adachi}}, \bibinfo
  {author} {\bibfnamefont {K.}~\bibnamefont {Takasan}}, \bibinfo {author}
  {\bibfnamefont {M.}~\bibnamefont {Ueda}},\ and\ \bibinfo {author}
  {\bibfnamefont {N.}~\bibnamefont {Kawakami}},\ }\bibfield  {title} {\bibinfo
  {title} {Theory of non-hermitian fermionic superfluidity with a
  complex-valued interaction},\ }\href
  {https://doi.org/10.1103/PhysRevLett.123.123601} {\bibfield  {journal}
  {\bibinfo  {journal} {Phys. Rev. Lett.}\ }\textbf {\bibinfo {volume} {123}},\
  \bibinfo {pages} {123601} (\bibinfo {year} {2019})}\BibitemShut {NoStop}%
\bibitem [{\citenamefont {Yamamoto}\ \emph {et~al.}(2021)\citenamefont
  {Yamamoto}, \citenamefont {Nakagawa}, \citenamefont {Tsuji}, \citenamefont
  {Ueda},\ and\ \citenamefont {Kawakami}}]{Kazuki_2021}%
  \BibitemOpen
  \bibfield  {author} {\bibinfo {author} {\bibfnamefont {K.}~\bibnamefont
  {Yamamoto}}, \bibinfo {author} {\bibfnamefont {M.}~\bibnamefont {Nakagawa}},
  \bibinfo {author} {\bibfnamefont {N.}~\bibnamefont {Tsuji}}, \bibinfo
  {author} {\bibfnamefont {M.}~\bibnamefont {Ueda}},\ and\ \bibinfo {author}
  {\bibfnamefont {N.}~\bibnamefont {Kawakami}},\ }\bibfield  {title} {\bibinfo
  {title} {Collective excitations and nonequilibrium phase transition in
  dissipative fermionic superfluids},\ }\href
  {https://doi.org/10.1103/PhysRevLett.127.055301} {\bibfield  {journal}
  {\bibinfo  {journal} {Phys. Rev. Lett.}\ }\textbf {\bibinfo {volume} {127}},\
  \bibinfo {pages} {055301} (\bibinfo {year} {2021})}\BibitemShut {NoStop}%
\bibitem [{\citenamefont {Scazza}\ \emph {et~al.}(2014)\citenamefont {Scazza},
  \citenamefont {Hofrichter}, \citenamefont {H{\"o}fer}, \citenamefont
  {De~Groot}, \citenamefont {Bloch},\ and\ \citenamefont
  {F{\"o}lling}}]{Scazza_2014}%
  \BibitemOpen
  \bibfield  {author} {\bibinfo {author} {\bibfnamefont {F.}~\bibnamefont
  {Scazza}}, \bibinfo {author} {\bibfnamefont {C.}~\bibnamefont {Hofrichter}},
  \bibinfo {author} {\bibfnamefont {M.}~\bibnamefont {H{\"o}fer}}, \bibinfo
  {author} {\bibfnamefont {P.~C.}\ \bibnamefont {De~Groot}}, \bibinfo {author}
  {\bibfnamefont {I.}~\bibnamefont {Bloch}},\ and\ \bibinfo {author}
  {\bibfnamefont {S.}~\bibnamefont {F{\"o}lling}},\ }\bibfield  {title}
  {\bibinfo {title} {Observation of two-orbital spin-exchange interactions with
  ultracold su(n)-symmetric fermions},\ }\href
  {https://doi.org/10.1038/nphys3061} {\bibfield  {journal} {\bibinfo
  {journal} {Nature Physics}\ }\textbf {\bibinfo {volume} {10}},\ \bibinfo
  {pages} {779} (\bibinfo {year} {2014})}\BibitemShut {NoStop}%
\bibitem [{\citenamefont {Pagano}\ \emph {et~al.}(2014)\citenamefont {Pagano},
  \citenamefont {Mancini}, \citenamefont {Cappellini}, \citenamefont
  {Lombardi}, \citenamefont {Sch{\"a}fer}, \citenamefont {Hu}, \citenamefont
  {Liu}, \citenamefont {Catani}, \citenamefont {Sias}, \citenamefont
  {Inguscio},\ and\ \citenamefont {Fallani}}]{Pagano_2014}%
  \BibitemOpen
  \bibfield  {author} {\bibinfo {author} {\bibfnamefont {G.}~\bibnamefont
  {Pagano}}, \bibinfo {author} {\bibfnamefont {M.}~\bibnamefont {Mancini}},
  \bibinfo {author} {\bibfnamefont {G.}~\bibnamefont {Cappellini}}, \bibinfo
  {author} {\bibfnamefont {P.}~\bibnamefont {Lombardi}}, \bibinfo {author}
  {\bibfnamefont {F.}~\bibnamefont {Sch{\"a}fer}}, \bibinfo {author}
  {\bibfnamefont {H.}~\bibnamefont {Hu}}, \bibinfo {author} {\bibfnamefont
  {X.-J.}\ \bibnamefont {Liu}}, \bibinfo {author} {\bibfnamefont
  {J.}~\bibnamefont {Catani}}, \bibinfo {author} {\bibfnamefont
  {C.}~\bibnamefont {Sias}}, \bibinfo {author} {\bibfnamefont {M.}~\bibnamefont
  {Inguscio}},\ and\ \bibinfo {author} {\bibfnamefont {L.}~\bibnamefont
  {Fallani}},\ }\bibfield  {title} {\bibinfo {title} {A one-dimensional liquid
  of fermions with tunable spin},\ }\href {https://doi.org/10.1038/nphys2878}
  {\bibfield  {journal} {\bibinfo  {journal} {Nature Physics}\ }\textbf
  {\bibinfo {volume} {10}},\ \bibinfo {pages} {198} (\bibinfo {year}
  {2014})}\BibitemShut {NoStop}%
\bibitem [{\citenamefont {Franchi}\ \emph {et~al.}(2017)\citenamefont
  {Franchi}, \citenamefont {Livi}, \citenamefont {Cappellini}, \citenamefont
  {Binella}, \citenamefont {Inguscio}, \citenamefont {Catani},\ and\
  \citenamefont {Fallani}}]{Franchi_2017}%
  \BibitemOpen
  \bibfield  {author} {\bibinfo {author} {\bibfnamefont {L.}~\bibnamefont
  {Franchi}}, \bibinfo {author} {\bibfnamefont {L.~F.}\ \bibnamefont {Livi}},
  \bibinfo {author} {\bibfnamefont {G.}~\bibnamefont {Cappellini}}, \bibinfo
  {author} {\bibfnamefont {G.}~\bibnamefont {Binella}}, \bibinfo {author}
  {\bibfnamefont {M.}~\bibnamefont {Inguscio}}, \bibinfo {author}
  {\bibfnamefont {J.}~\bibnamefont {Catani}},\ and\ \bibinfo {author}
  {\bibfnamefont {L.}~\bibnamefont {Fallani}},\ }\bibfield  {title} {\bibinfo
  {title} {State-dependent interactions in ultracold 174yb probed by optical
  clock spectroscopy},\ }\href {https://doi.org/10.1088/1367-2630/aa8fb4}
  {\bibfield  {journal} {\bibinfo  {journal} {New J. Phys.}\ }\textbf {\bibinfo
  {volume} {19}},\ \bibinfo {pages} {103037} (\bibinfo {year}
  {2017})}\BibitemShut {NoStop}%
\bibitem [{\citenamefont {Bouganne}\ \emph {et~al.}(2017)\citenamefont
  {Bouganne}, \citenamefont {Aguilera}, \citenamefont {Dareau}, \citenamefont
  {Soave}, \citenamefont {Beugnon},\ and\ \citenamefont
  {Gerbier}}]{Bouganne_2017}%
  \BibitemOpen
  \bibfield  {author} {\bibinfo {author} {\bibfnamefont {R.}~\bibnamefont
  {Bouganne}}, \bibinfo {author} {\bibfnamefont {M.~B.}\ \bibnamefont
  {Aguilera}}, \bibinfo {author} {\bibfnamefont {A.}~\bibnamefont {Dareau}},
  \bibinfo {author} {\bibfnamefont {E.}~\bibnamefont {Soave}}, \bibinfo
  {author} {\bibfnamefont {J.}~\bibnamefont {Beugnon}},\ and\ \bibinfo {author}
  {\bibfnamefont {F.}~\bibnamefont {Gerbier}},\ }\bibfield  {title} {\bibinfo
  {title} {Clock spectroscopy of interacting bosons in deep optical lattices},\
  }\href {https://doi.org/10.1088/1367-2630/aa8c45} {\bibfield  {journal}
  {\bibinfo  {journal} {New J. Phys.}\ }\textbf {\bibinfo {volume} {19}},\
  \bibinfo {pages} {113006} (\bibinfo {year} {2017})}\BibitemShut {NoStop}%
\bibitem [{\citenamefont {Gorshkov}\ \emph {et~al.}(2010)\citenamefont
  {Gorshkov}, \citenamefont {Hermele}, \citenamefont {Gurarie}, \citenamefont
  {Xu}, \citenamefont {Julienne}, \citenamefont {Ye}, \citenamefont {Zoller},
  \citenamefont {Demler}, \citenamefont {Lukin},\ and\ \citenamefont
  {Rey}}]{Gorshkov_2010}%
  \BibitemOpen
  \bibfield  {author} {\bibinfo {author} {\bibfnamefont {A.~V.}\ \bibnamefont
  {Gorshkov}}, \bibinfo {author} {\bibfnamefont {M.}~\bibnamefont {Hermele}},
  \bibinfo {author} {\bibfnamefont {V.}~\bibnamefont {Gurarie}}, \bibinfo
  {author} {\bibfnamefont {C.}~\bibnamefont {Xu}}, \bibinfo {author}
  {\bibfnamefont {P.~S.}\ \bibnamefont {Julienne}}, \bibinfo {author}
  {\bibfnamefont {J.}~\bibnamefont {Ye}}, \bibinfo {author} {\bibfnamefont
  {P.}~\bibnamefont {Zoller}}, \bibinfo {author} {\bibfnamefont
  {E.}~\bibnamefont {Demler}}, \bibinfo {author} {\bibfnamefont {M.~D.}\
  \bibnamefont {Lukin}},\ and\ \bibinfo {author} {\bibfnamefont {A.~M.}\
  \bibnamefont {Rey}},\ }\bibfield  {title} {\bibinfo {title} {Two-orbital s
  u(n) magnetism with ultracold alkaline-earth atoms},\ }\href
  {https://doi.org/10.1038/nphys1535} {\bibfield  {journal} {\bibinfo
  {journal} {Nature Physics}\ }\textbf {\bibinfo {volume} {6}},\ \bibinfo
  {pages} {289} (\bibinfo {year} {2010})}\BibitemShut {NoStop}%
\bibitem [{\citenamefont {Cazalilla}\ and\ \citenamefont
  {Rey}(2014)}]{Cazalilla_2014}%
  \BibitemOpen
  \bibfield  {author} {\bibinfo {author} {\bibfnamefont {M.~A.}\ \bibnamefont
  {Cazalilla}}\ and\ \bibinfo {author} {\bibfnamefont {A.~M.}\ \bibnamefont
  {Rey}},\ }\bibfield  {title} {\bibinfo {title} {Ultracold fermi gases with
  emergent su(n) symmetry},\ }\href
  {https://doi.org/10.1088/0034-4885/77/12/124401} {\bibfield  {journal}
  {\bibinfo  {journal} {Reports on Progress in Physics}\ }\textbf {\bibinfo
  {volume} {77}},\ \bibinfo {pages} {124401} (\bibinfo {year}
  {2014})}\BibitemShut {NoStop}%
\bibitem [{\citenamefont {Georgi}(1999)}]{GeorgiBook}%
  \BibitemOpen
  \bibfield  {author} {\bibinfo {author} {\bibfnamefont {H.}~\bibnamefont
  {Georgi}},\ }\href {https://doi.org/10.1201/9780429499210} {\emph {\bibinfo
  {title} {Lie Algebras in Particle Physics}}}\ (\bibinfo  {publisher} {CRC
  Press},\ \bibinfo {year} {1999})\BibitemShut {NoStop}%
\bibitem [{\citenamefont {Hartmann}(2016)}]{Hartmann_2016}%
  \BibitemOpen
  \bibfield  {author} {\bibinfo {author} {\bibfnamefont {S.}~\bibnamefont
  {Hartmann}},\ }\href@noop {} {\bibinfo {title} {Generalized dicke states}}
  (\bibinfo {year} {2016}),\ \Eprint {https://arxiv.org/abs/1201.1732}
  {arXiv:1201.1732 [quant-ph]} \BibitemShut {NoStop}%
\bibitem [{\citenamefont {Rosso}\ \emph {et~al.}()\citenamefont {Rosso},
  \citenamefont {Mazza},\ and\ \citenamefont {Biella}}]{SuppMat}%
  \BibitemOpen
  \bibfield  {author} {\bibinfo {author} {\bibfnamefont {L.}~\bibnamefont
  {Rosso}}, \bibinfo {author} {\bibfnamefont {L.}~\bibnamefont {Mazza}},\ and\
  \bibinfo {author} {\bibfnamefont {A.}~\bibnamefont {Biella}},\ }\bibfield
  {title} {\bibinfo {title} {Supplemental material},\ }\href@noop {} {\bibinfo
  {journal} {URL will be inserted by the publisher}\ }\BibitemShut {NoStop}%
\bibitem [{Note1()}]{Note1}%
  \BibitemOpen
\bibfield  {journal} {  }\bibinfo {note} {Taking the thermodynamic limit of
  Eq.~\protect \textup {\hbox {\mathsurround \z@ \protect \normalfont
  (\ignorespaces \ref {dickecondition}\unskip \@@italiccorr )}} we get $$
  \protect \qopname \relax m{lim}_{L\to \infty }\protect \frac {\delimiter
  "426830A \protect \mathaccentV {hat}05ES^{2}_{AB}\delimiter "526930B }{L^{2}}
  = \protect \frac {\hbar ^{2}}{4} \protect \frac {\delimiter "426830A \protect
  \mathaccentV {hat}05EN_{AB}^{2}\delimiter "526930B }{L^{2}}, $$ which gives
  the relation~\protect \textup {\hbox {\mathsurround \z@ \protect \normalfont
  (\ignorespaces \ref {dickesu2}\unskip \@@italiccorr )}}.}\BibitemShut {Stop}%
\bibitem [{\citenamefont {Johansson}\ \emph {et~al.}(2012)\citenamefont
  {Johansson}, \citenamefont {Nation},\ and\ \citenamefont {Nori}}]{Qutip01}%
  \BibitemOpen
  \bibfield  {author} {\bibinfo {author} {\bibfnamefont {J.}~\bibnamefont
  {Johansson}}, \bibinfo {author} {\bibfnamefont {P.}~\bibnamefont {Nation}},\
  and\ \bibinfo {author} {\bibfnamefont {F.}~\bibnamefont {Nori}},\ }\bibfield
  {title} {\bibinfo {title} {Qutip: An open-source python framework for the
  dynamics of open quantum systems},\ }\href
  {https://doi.org/https://doi.org/10.1016/j.cpc.2012.02.021} {\bibfield
  {journal} {\bibinfo  {journal} {Computer Physics Communications}\ }\textbf
  {\bibinfo {volume} {183}},\ \bibinfo {pages} {1760} (\bibinfo {year}
  {2012})}\BibitemShut {NoStop}%
\bibitem [{\citenamefont {Johansson}\ \emph {et~al.}(2013)\citenamefont
  {Johansson}, \citenamefont {Nation},\ and\ \citenamefont {Nori}}]{Qutip02}%
  \BibitemOpen
  \bibfield  {author} {\bibinfo {author} {\bibfnamefont {J.}~\bibnamefont
  {Johansson}}, \bibinfo {author} {\bibfnamefont {P.}~\bibnamefont {Nation}},\
  and\ \bibinfo {author} {\bibfnamefont {F.}~\bibnamefont {Nori}},\ }\bibfield
  {title} {\bibinfo {title} {Qutip 2: A python framework for the dynamics of
  open quantum systems},\ }\href
  {https://doi.org/https://doi.org/10.1016/j.cpc.2012.11.019} {\bibfield
  {journal} {\bibinfo  {journal} {Computer Physics Communications}\ }\textbf
  {\bibinfo {volume} {184}},\ \bibinfo {pages} {1234} (\bibinfo {year}
  {2013})}\BibitemShut {NoStop}%
\end{thebibliography}%

\clearpage
\onecolumngrid 
\appendix
\begin{center}
\textbf{\large Supplemental Material} \\
\vspace{+0.4em}
{\large The eightfold way to dark states in SU($3$) cold gases with two-body losses} \\    
\vspace{+0.3em}
{ Lorenzo Rosso$^{1}$, Leonardo Mazza$^{1}$ and Alberto Biella$^{2.1}$}\\
\vspace{+0.2em}
{\small \itshape $^{1}$Universit\'e Paris-Saclay, CNRS, LPTMS, 91405 Orsay, France}\\
{\small \itshape $^{2}$INO-CNR BEC Center and Dipartimento di Fisica, Universit\`a di Trento, 38123 Povo, Italy}

\end{center}
\section{I. Populations dynamics}
\label{appmain}
The spin-resolved populations obey the following equation (see Eq.~(5) in the main text)
\begin{equation}
\label{popapp}
\dot N_{\mu}(t) = -\gamma \sum_{j}\sum_{\mu\neq\mu'}\Big\langle\hat n_{j,\mu} \hat n_{j,\mu'}\Big\rangle.
\end{equation} 

\begin{proof}
From the Lindblad equation (see Eq.~(4) in the main text) and using the ciclic property of the trace we get
\begin{eqnarray}
\label{proof1}
 \dot N_{\mu}(t) &=&   \sum_\alpha \left(\Big\langle \hat{L}^{\dagger}_{\alpha}\hat N_\mu \hat{L}_{\alpha}\Big\rangle_t -\frac12\Big\langle\left\{ \hat L_\alpha^\dagger \hat L_\alpha, \hat N_{\mu} \right\}\Big\rangle_t\right)\cr
 &=& \sum_\alpha \left(\Big\langle \hat{L}^{\dagger}_{\alpha}\hat N_{\mu}\hat{L}_{\alpha}\Big\rangle_t -\Big\langle \hat L_\alpha^\dagger \hat L_\alpha \hat N_{\mu} \Big\rangle_t\right)\cr
 &=& \sum_\alpha \Big\langle \hat{L}^{\dagger}_{\alpha}\left[\hat N_{\mu},\hat{L}_{\alpha}\right]\Big\rangle_t, 
\end{eqnarray}
where in the second line we used that 
\begin{eqnarray}
\left[\hat L_j^\dagger \hat L_j,\hat N_{\mu}\right]&=&\left[ (\hat L_{j}^{\sigma\sigma'})^{\dagger} \hat L_{j}^{\sigma\sigma'},\hat N_{\mu}\right] \cr
&=&\gamma \  \left[\hat c_{j, \sigma'}^{\dagger} \hat c_{j, \sigma}^{\dagger} \hat c_{j, \sigma} \hat c_{j, \sigma'}, \hat N_{\mu}\right] \cr
&=&\gamma \  \left[\hat N_{j,\sigma}\hat N_{j,\sigma'}, \hat N_{\mu}\right] \cr
&=&0.
\end{eqnarray}
Let us now evaluate the commutator $\left[\hat N_{\mu},\hat L_{j}^{\sigma\sigma'}\right]$.
This term non-zero only if $\sigma=\mu$ or $\sigma'=\mu$. 
We define the a generic $N$-particle state $\ket{\{N_{\eta}\}}$ and compute
\begin{eqnarray}
\hat N_{\mu}\hat{L}_{j}^{\sigma\sigma'} \ket{\{N_{\eta}\}}&=&  \hat N_{\mu} \sqrt{\gamma} \  \hat c_{j, \mu} \hat c_{j, \sigma'} \ket{\{N_{\eta}\}}\cr
&&\cr
&=&  \hat N_{\mu} \sqrt{\gamma} \ \ket{\{N_{\eta}\}_{\eta\neq{\sigma,\sigma'}},  N_{\sigma-1},  N_{\sigma'-1}}\cr
&&\cr
&=&  (N_{\mu}-\delta_{\sigma\mu}-\delta_{\sigma'\mu}) \hat{L}_{j}^{\sigma\sigma'} \ket{\{N_{\eta}\}}.
\end{eqnarray}
We also get 
\begin{equation}
\hat{L}_{j}^{\sigma\sigma'} \hat N_{\mu} \ket{\{N_{\eta}\}}=N_{\mu} \hat{L}_{j}^{\sigma\sigma'}\ket{\{N_{\eta}\}}.
\end{equation}
So that 
\begin{equation}
\label{comm}
\left[\hat N_{\mu},\hat{L}_{j}^{\sigma\sigma'}\right] =- \hat{L}_{j}^{\sigma\sigma'} \left(\delta_{\sigma\mu}+ \delta_{\sigma'\mu}\right).
\end{equation}
Thus, inserting Eq.~\eqref{comm} in Eq.~\eqref{proof1} we get Eq.~\eqref{popapp}.
\end{proof}
From Eq.~\eqref{comm} we also easily get 
\begin{eqnarray}
2\left[\hat\Lambda^{0}_{\mu\mu'}, \hat{L}_{j}^{\sigma\sigma'}\right] =  -\hat{L}_{j}
^{\sigma\sigma'} \left(\delta_{\sigma\mu}+ \delta_{\sigma'\mu}+\delta_{\sigma\mu'}+ \delta_{\sigma'\mu'} \right), \cr
&&\cr
2\left[\hat\Lambda^{z}_{\mu\mu'}, \hat{L}_{j}^{\sigma\sigma'}\right] =  -\hat{L}_{j}
^{\sigma\sigma'}\left(\delta_{\sigma\mu}+\delta_{\sigma'\mu}-\delta_{\sigma\mu'}- \delta_{\sigma'\mu'} \right).\cr
\end{eqnarray}

\section{II. Proof of the generalised Dicke state relation}

The goal of this section is to prove that any state belonging to a representation $(p,0)$ of the SU(3) group satisfies:
\begin{equation}
  \frac{1}{\hbar^2}\langle \hat S_{\mu \mu'}^2 \rangle = \left\langle \frac{\hat N_{\mu,\mu'}}{2} \left( \frac{\hat N_{\mu,\mu'}}{2}+1\right) \right\rangle .
  \label{GDS:STATEMENT}
\end{equation}

Let us consider the diagram in Fig.~1 of the main text, where we focused on the ten states of the representation $(3,0)$. We redraw it focusing 
on the observables $\Lambda^{0}_{AB}=(\hat n_A+ \hat n_B)/2$ and $\Lambda_{AB}^z = (\hat n_A-\hat n_B)/2$. 
The axes cross at the origin, in correspondence with the state with $\Lambda_{AB}^0=0$ and $\Lambda_{AB}^z=0$.

\begin{figure}[h]
\includegraphics{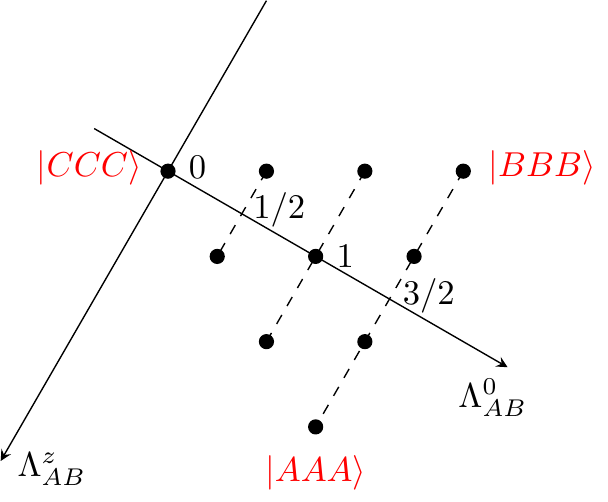}
\end{figure}

By simple observation, it is easy to establish that we have one state such that $\Lambda^0_{AB}=0$, two states such that $\Lambda^0_{AB}=1/2$, three states such that $\Lambda^{0}_{AB}=1$ and in general that the number of states at fixed $\Lambda^{0}_{AB}$ is $2\Lambda^{0}_{AB}+1$. These multiplets are highlighted by the dashed lines at fixed $\Lambda^0_{AB}$, whose value is indicated by the number in black.

Since we can use the operators $\Lambda_{AB}^{\alpha}$ with $\alpha = x,y,z$ to define three SU(2) spin operators: $S^{\alpha}_{AB} = \hbar\Lambda^{\alpha}_{AB}$, we then have that at fixed $\Lambda^0_{AB}$ the spin operator $S^z_{AB}$ takes values between $-\Lambda_{AB}^0$ and $+\Lambda_{AB}^0$ at integer steps. The states thus belong to a representation of $\hat S^2_{AB} = (\hat S^x_{AB})^{2}+(\hat S^y_{AB})^{2}+(\hat S^z_{AB})^{2}$ with quantum number $\Lambda^0_{AB}$, hence the thesis in Eq.~\eqref{GDS:STATEMENT} for $\mu \mu'=AB $.

The reasoning can be generalised also to the other two pseudo-spins $\Lambda_{AC}$ and $\Lambda_{BC}$, so that the statement in Eq.~\eqref{GDS:STATEMENT} is true in full generality for any pair of indexes $\mu \mu'$.

\section{III. Details about the weakly dissipative and weakly interacting case}
\label{correlatorsweak}
In this regime it is useful to expand Eq.~(5) (main text) on the basis of plane waves $\hat{c}_{k,\sigma}=L^{-1/2}\sum_{j}e^{i k j}\hat{c}_{kj,\sigma}$. We get
\begin{equation}
\label{pw}
\dot N_{\mu}(t) = - \frac{\gamma}{L} \sum_{k,q,w,z} \sum_{n}\sum_{\sigma\neq\mu} \Big\langle  \hat c_{k, \mu}^{\dagger} \hat c_{w, \mu} \hat c_{q, \sigma}^{\dagger} \hat c_{z, \sigma} \Big\rangle_t \ \delta_{k+q-w-z,2\pi n},
\end{equation}
where  the Kronecker delta ensures the conservation of the momentum (modulus $2\pi$).
The Hamiltonian time evolution of the correlators appearing in Eq.~\eqref{pw} can be written as 
\begin{equation}
 \label{pw2}
\langle  \hat c_{k, \mu}^{\dagger} \hat c_{w, \mu} \hat c_{q, \sigma}^{\dagger} \hat c_{z, \sigma} \rangle_t = e^{-\frac{i}{\hbar}(E_{k}+E_{q}-E_{w}-E_{z})t}\langle  \hat c_{k, \mu}^{\dagger} \hat c_{w, \mu} \hat c_{q, \sigma}^{\dagger} \hat c_{z, \sigma} \rangle_0,
\end{equation}
where $E_{k}=-2J \cos(k)$ is the energy of the eigenstate with quasi-momentum $k$ of the free-fermion Hamiltonian.
In analogy to the SU($2$) case studied in Ref.~\cite{Rosso_2021} we keep only the energy-conserving correlators.
Within this approximation we get
\begin{equation}
\label{eqweak1}
 \dot N_{\mu}(t) = - \frac{\gamma}{L} \sum_{\mu'\neq\mu}\left[\frac{\langle\hat N_{\mu\mu'}^{2}\rangle_{t}}{4} + \frac{\langle\hat N_{\mu\mu'}\rangle_{t}}{2} - \frac{\langle\hat S_{\mu\mu'}^{2}\rangle_{t}}{\hbar^{2}} + \langle\hat C_{\mu\mu'}\rangle_{t} \right],
\end{equation}  
where $\hat N_{\mu\mu'}=\hat N_{\mu} + \hat N_{\mu'}= 2 \hat \Lambda^{0}_{\mu\mu'}$.
The operator $\hat C_{\mu\mu'}$ accounts for different kind of correlations between the $\mu$ and $\mu'$ spin sectors that will not be relevant in the thermodynamic limit $L\to\infty$ and has the form 
\begin{equation}
\hat C_{\sigma\mu}=- \langle\hat \Pi_{\sigma\mu}\rangle_{t} + \langle\hat \Sigma_{\sigma\mu}\rangle_{t} +\langle\hat T_{\sigma\mu}\rangle_{t},
\end{equation}
where 
\begin{eqnarray}
\hat\Pi_{\sigma\mu} &=& \sum_{k}\hat n_{k,\sigma}\hat n_{k,\mu},\cr
\hat \Sigma_{\sigma\mu} &=& \sum_{k \neq q, \; k \neq \pi-q} 
    \hat c_{k, \sigma}^\dagger 
    \hat c_{q, \sigma}
    \hat c_{\pi- k, \mu}^\dagger 
    \hat c_{\pi-q, \mu},  \\
\hat T_{\sigma\mu} &=&  \sum_{\delta k \in \left[0, \frac \pi 2 \right]} \left( 
    \hat c_{\frac \pi 2 + \delta k, \sigma}^\dagger \hat c_{- \frac \pi 2 - \delta k, \sigma} \hat c_{ \frac \pi 2-\delta k, \mu}^\dagger \hat c_{- \frac \pi 2 + \delta k, \mu}
    + 
    \hat c_{\frac \pi 2 - \delta k, \sigma}^\dagger \hat c_{- \frac \pi 2 - \delta k, \sigma} \hat c_{ \frac \pi 2+\delta k, \mu}^\dagger \hat c_{- \frac \pi 2 + \delta k, \mu}
    +{\rm H.c.}\right).
\end{eqnarray}
The operator $\hat \Pi_{\sigma\mu}$ is a density-density correlator between the $\sigma$ and the $\mu$ spin sectors, $\hat \Sigma_{\sigma\mu}$ accounts for correlators that are symmetric with respect to the center of the band, located at $k = \pm \pi/2$ (note that in this operator momenta are defined mod $2 \pi$ to  restrict them to the first Brillouin zone), and $\hat T_u$ considers umklapp terms, where the difference in momenta is equal to $\pm 2 \pi$. 

\section{IV. Dynamics of coherences and Gaussian approximation}
\label{othercomm}
Starting again from the Lindblad master equation~(4) (main text) we get 
\begin{equation}
\label{motionxy}
 \dot \Lambda_{\mu\mu'}^{x,y}(t) =  \sum_\alpha \Big\langle \hat{L}^{\dagger}_{\alpha}\left[\hat \Lambda_{\mu\mu'}^{x,y},\hat{L}_{\alpha}\right]\Big\rangle_t,
\end{equation}
where $\Lambda_{\mu\mu'}^{x,y}(t)=\langle \hat \Lambda_{\mu\mu'}^{x,y}\rangle_{t}$ and we used the fact that $\left[\hat L_j^\dagger \hat L_j, \hat \Lambda_{\mu\mu'}^{x,y}\right]=0.$
Using the defintion~\eqref{gensun}, after some algebra we get
\begin{eqnarray}
2 \left[\hat\Lambda^{x}_{\mu\mu'}, \hat{L}_{j}^{\sigma\sigma'}\right] &=&  
-\hat{L}_{j}^{\mu'\sigma'} \delta_{\sigma\mu} 
-\hat{L}_{j} ^{\sigma\mu'} \delta_{\mu\sigma'} 
-\hat{L}_{j}^{\mu\sigma'} \delta_{\sigma\mu'}
-\hat{L}_{j}^{\sigma\mu} \delta_{\sigma'\mu'}, \cr
&&\cr
2 \left[\hat\Lambda^{y}_{\mu\mu'}, \hat{L}_{j}^{\sigma\sigma'}\right] &=&  -i\big(
-\hat{L}_{j}^{\mu'\sigma'} \delta_{\sigma\mu} 
-\hat{L}_{j} ^{\sigma\mu'} \delta_{\mu\sigma'} 
+\hat{L}_{j}^{\mu\sigma'} \delta_{\sigma\mu'}
+\hat{L}_{j}^{\sigma\mu} \delta_{\sigma'\mu'}\big),
\end{eqnarray}
and thus for $\hat\Lambda^{\pm}_{\mu\mu'}=\hat\Lambda^{x}_{\mu\mu'}\pm i \hat\Lambda^{y}_{\mu\mu'}$ we obtain
\begin{eqnarray}
\label{commspm}
\left[\hat\Lambda^{+}_{\mu\mu'}, \hat{L}_{j}^{\sigma\sigma'}\right] &=&  -\big(
\hat{L}_{j}^{\mu'\sigma'} \delta_{\sigma\mu} 
+\hat{L}_{j} ^{\sigma\mu'} \delta_{\mu\sigma'}\big), \cr
&&\cr
\left[\hat\Lambda^{-}_{\mu\mu'}, \hat{L}_{j}^{\sigma\sigma'}\right] &=&  -\big(
\hat{L}_{j}^{\mu\sigma'} \delta_{\sigma\mu'}
+\hat{L}_{j}^{\sigma\mu} \delta_{\sigma'\mu'}\big).
\end{eqnarray}
Finally, combining~\eqref{commspm} and~\eqref{motionxy} we obtain Eq.~\eqref{coheeq}
\begin{eqnarray}
\dot\Lambda^{x}_{\mu\mu'}+i \dot\Lambda^{y}_{\mu\mu'}
&=&-\sum_{j}\sum_{\sigma\neq\mu,\mu'} \langle (\hat{L}_{j}
^{\mu\sigma})^{\dagger} \hat{L}_{j}
^{\mu'\sigma} \rangle \cr
 \dot\Lambda^{x}_{\mu\mu'}-i \dot\Lambda^{y}_{\mu\mu'}
&=&-\sum_{j}\sum_{\sigma\neq\mu,\mu'} \langle (\hat{L}_{j}
^{\mu'\sigma})^{\dagger} \hat{L}_{j}
^{\mu\sigma} \rangle,
\end{eqnarray}
in agreement with the fact that $ \hat\Lambda^{x}_{\mu\mu'}+i \hat\Lambda^{y}_{\mu\mu'}= \hat\Lambda^{x}_{\mu'\mu}-i \hat\Lambda^{y}_{\mu'\mu}$.
Let us now take the thermodynamic limit $L\to\infty$ of the above set of equations.
We divide both the sides by $L$ and introduce the intensive quantities $s^{\pm}_{\mu\mu'}(t)\equiv \langle\hat\Lambda^{x}_{\mu\mu'}\pm i \hat\Lambda^{y}_{\mu\mu'}\rangle_{t}/L$. For the $+$ coherence we get the following
\begin{eqnarray}
\label{gauss+}
\dot s^{+}_{\mu,\mu'}&=& -\frac{\gamma}{L}\sum_{j}\sum_{\sigma\neq\mu\mu'}\langle \hat c_{j,\sigma}^{\dagger} \hat c_{j,\mu}^{\dagger}  \hat c_{j,\mu'} \hat c_{j,\sigma}\rangle_{t} \cr
&=&-\frac{\gamma}{L^{3}}\sum_{j}\sum_{\sigma\neq\mu\mu'}\sum_{k,q,w,z} e^{i(k+q-w-z)j} \langle \hat c_{k,\sigma}^{\dagger} \hat c_{q,\mu}^{\dagger}  \hat c_{w,\mu'} \hat c_{z,\sigma}\rangle_{t}\cr
&=&-\frac{\gamma}{L^{2}}\sum_{\sigma\neq\mu\mu'}\sum_{k,q,w,z} \delta_{k+w,w+z} \langle \hat c_{k,\sigma}^{\dagger} \hat c_{q,\mu}^{\dagger}  \hat c_{w,\mu'} \hat c_{z,\sigma}\rangle_{t}.
\end{eqnarray}
Let us now assume that the density matrix is gaussian and exploit the Wick's theorem
\begin{equation}
\label{wick+}
 \langle \hat c_{k,\sigma}^{\dagger} \hat c_{q,\mu}^{\dagger}  \hat c_{w,\mu'} \hat c_{z,\sigma}\rangle_{t} \sim 
 \langle \hat c^{\dagger}_{k,\sigma} \hat c_{z,\sigma}\rangle_{t}
\langle \hat c^{\dagger}_{q,\mu} \hat c_{w,\mu'}\rangle_{t} \ \delta_{k,z}\  \delta_{q,w}
 -\langle \hat c^{\dagger}_{k,\sigma} \hat c_{w,\mu'} \rangle_{t} 
 \langle \hat c^{\dagger}_{q,\mu} \hat c_{z,\sigma}\rangle_{t}\ \delta_{k,w} \ \delta_{q,z},
\end{equation}
where the Kronecker delta selects only the two-point correlator which do not have an explicit time dependence. 
As for the populations this comes from the fact that dissipation is weak and time-dependent correlators will average to zero between two dissipative events.

Inserting Eq.~\eqref{wick+} into Eq.~\eqref{gauss+} se fnally get 
\begin{equation}
\label{coheeq}
\dot s^{\pm}_{\mu\mu'}=-\gamma \sum_{\sigma\neq\mu,\mu'}\big(
n_{\sigma} s^{\pm}_{\mu\mu'} - s^{\pm}_{\sigma\mu'}s^{\pm}_{\mu\sigma}\big).
\end{equation}

The set of equations~(7) (main text) can be easily recasted in a set of equations for
$s^{0,z}_{\mu\mu'}$. We obtain
\begin{equation}
\label{popgauss1}
\dot s^{0,z}_{\mu\mu'} = \frac{\gamma}{2}\sum_{\sigma\neq\eta}
 \vec{s}^{ {\ \mathsf T}}_{\sigma\eta} \ \mathsf{G} \ \vec{s}_{\sigma\eta} \ (\delta_{\eta\mu} \pm \delta_{\eta\mu'})
\end{equation}

Eq.~\eqref{coheeq} and~\eqref{popgauss1} are a closed set of equations for the four components of $\vec{s}_{\sigma\mu}$. This result allows to compute the time evolution of
any initial state within the Gaussian approximation.

\section{V. Perturbative solutions in the weakly-dissipative regime}
\begin{figure}[t]
\includegraphics[width=0.35\columnwidth]{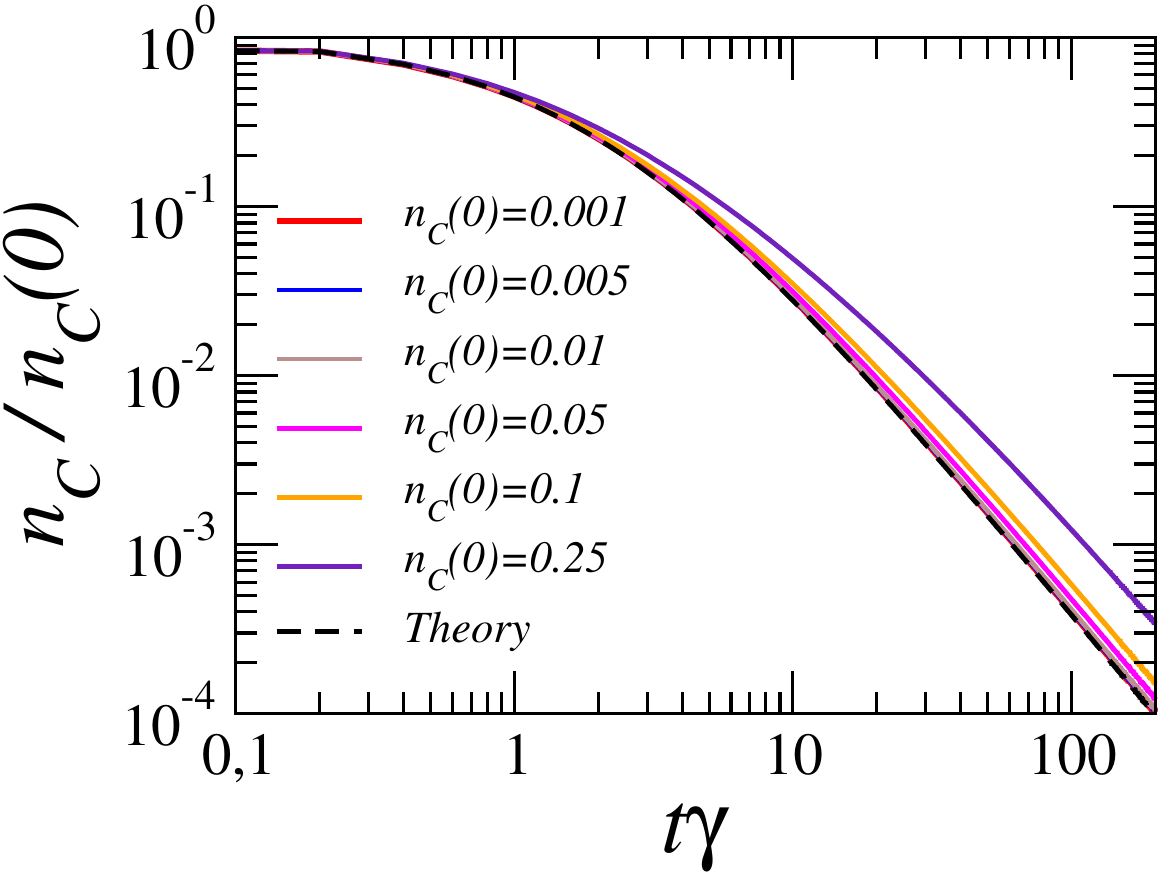}
\caption{SU($3$) dynamics in the weakly dissipative limit. We compare the behavior of $n_{C}(t)/n_{C}(0)$ for different values of $n_{C}(0)=\lambda$ with the perturbative result~\eqref{appwe04}. Here we set $n_{A}(t)=0.8$.}
\label{Fig:weaksol}
\end{figure}

In this section we derive some solution for the $N=3$ case in the weakly-dissipative regime and in absence of coherences $s^{x,y}_{\sigma,\sigma'}=0$ for $\sigma,\sigma'=A,B,C$ and $\sigma<\sigma'$.
The dynamics is ruled by Eq.~(9) (main text) that for $N=3$ gives 
\begin{eqnarray}
\label{appwe01}
\dot{n}_{A} &=& -\gamma n_{A} (n_{B}+n_{C}), \cr
\dot{n}_{B} &=& -\gamma n_{B} (n_{A}+n_{C}), \cr
\dot{n}_{C} &=& -\gamma n_{C} (n_{A}+n_{B}). 
\end{eqnarray}
We now consider the case where the population in $A$ and $B$ sector is the same $n_{A}(t)=n_{B}(t)$ and the system is initially prepared with a large fraction of the population in $A,B$  and a small amount of population in the $C$ sector, i.e $n_{A}(0)=n_{B}(0)\gg n_{C}(0)=\lambda$.
We can thus use the following Taylor expansion in the small parameter $\lambda$ 
\begin{eqnarray}
\label{appwe02}
n_{A} (t) &=& n_{A}^{(0)} (t) + \lambda n_{A}^{(1)} (t) + \mathcal{O} (\lambda^{2}),\cr
n_{C}(t) &=& \lambda n_{C}^{(1)} (t) + \mathcal{O} (\lambda^{2}),
\end{eqnarray}
and the initial conditions translates into $n_{C}^{(1)} (0)=1$ and $n (0)=n^{(0)} (0)$.
Inserting Eq.~\eqref{appwe02} into Eq.~\eqref{appwe01} we get at first order in $\lambda$
\begin{eqnarray}
\label{appwe03}
\dot{n}_{A}^{(0)} &=& -\gamma \left(n_{A}^{(0)} \right)^{2}, \cr
\dot{n}_{A}^{(1)}&=&  -\gamma n_{A}^{(0)} \left(2 n_{A}^{(1)} + n_{C}^{(1)}\right), \cr
\dot{n}^{(1)}_{C}&=&-2\gamma n_{C}^{(1)} n_{A}^{(0)}.
\end{eqnarray}
The set of equations~\eqref{appwe03} can be solved exactly. We get 
\begin{equation}
\label{appwe04}
n_{A}^{(0)} (t) = \frac{n_{A}^{(0)} (0) }{1+\gamma t  \ n_{A}^{(0)} (0)}, \quad
n_{A}^{(1)} =  -\frac{\ln[1+\gamma t  \ n_{A}^{(0)} (0))]}{\left[ 1+\gamma t  \ n_{A}^{(0)} (0)\right]^{2}}, \quad
n^{(1)}_{C} = \frac{1}{\left[ 1+\gamma t  \ n_{A}^{(0)} (0)\right]^{2}}.
\end{equation}
The solution~\eqref{appwe04} predicts that the systems gets empty in the long time limit, i.e.
$
\lim_{t\to\infty} n(t) = \lim_{t\to\infty} n_{C}(t) =0.
$
The perturbative result~\eqref{appwe04} obtained for $n_{C}(t)$ is in very good agreement with the exact numerical integration of Eq.~\eqref{appwe02} as shown in Fig.~\ref{Fig:weaksol}.

We now derive an approximate solution when the system is initially prepared with a large fraction of the total population in the $A$ sector and a small (and equal) fraction of particles in the $B,C$ sectors, i.e. $n_{B}(0)=n_{C}(0)=\lambda\ll n_{A}(0)$. During the dynamics the population in the $B,C$ sectors remain equal $n_{B}(t)=n_{C}(t)$ and, 
as we did before, we can exploit a Taylor expansion for the population densities
\begin{eqnarray}
\label{appwe05}
n_{A} (t) &=& n_{A}^{(0)} (t) + \lambda n^{(1)}_{A} (t) + \mathcal{O} (\lambda^{2}),\cr
n_{C}(t) &=& \lambda n_{C}^{(1)} (t) + \mathcal{O} (\lambda^{2}),
\end{eqnarray} 
where $n_{C}^{(1)}(0)=1$ and $n_{A} (0)=n_{A}^{(0)} (0)$.
Inserting Eq.~\eqref{appwe05} into Eq.~\eqref{appwe01} we get 
\begin{eqnarray}
\label{appwe06}
\dot n_{A}^{(0)} (t) &=& 0, \cr
\dot n_{A}^{(1)} (t) &=& -2\gamma n_{A}^{(0)} n_{C}^{(1)}, \cr
\dot n_{C}^{(1)} (t) &=& -\gamma n_{A}^{(0)} n_{C}^{(1)}.
\end{eqnarray}
The set of equations~\eqref{appwe06} can be solved exactly. We get 
\begin{equation}
\label{appwe07}
n_{A}^{(0)} (t) = n_{A}^{(0)} (0), \quad
n_{A}^{(1)} =  -2 \left(1-e^{-\gamma n_{A}^{(0)}(0)} \right), \quad
n_{C}^{(1)} = e^{-\gamma n_{A}^{(0)}(0) t}.
\end{equation}
\begin{figure}[t!]
\includegraphics[width=0.35\columnwidth]{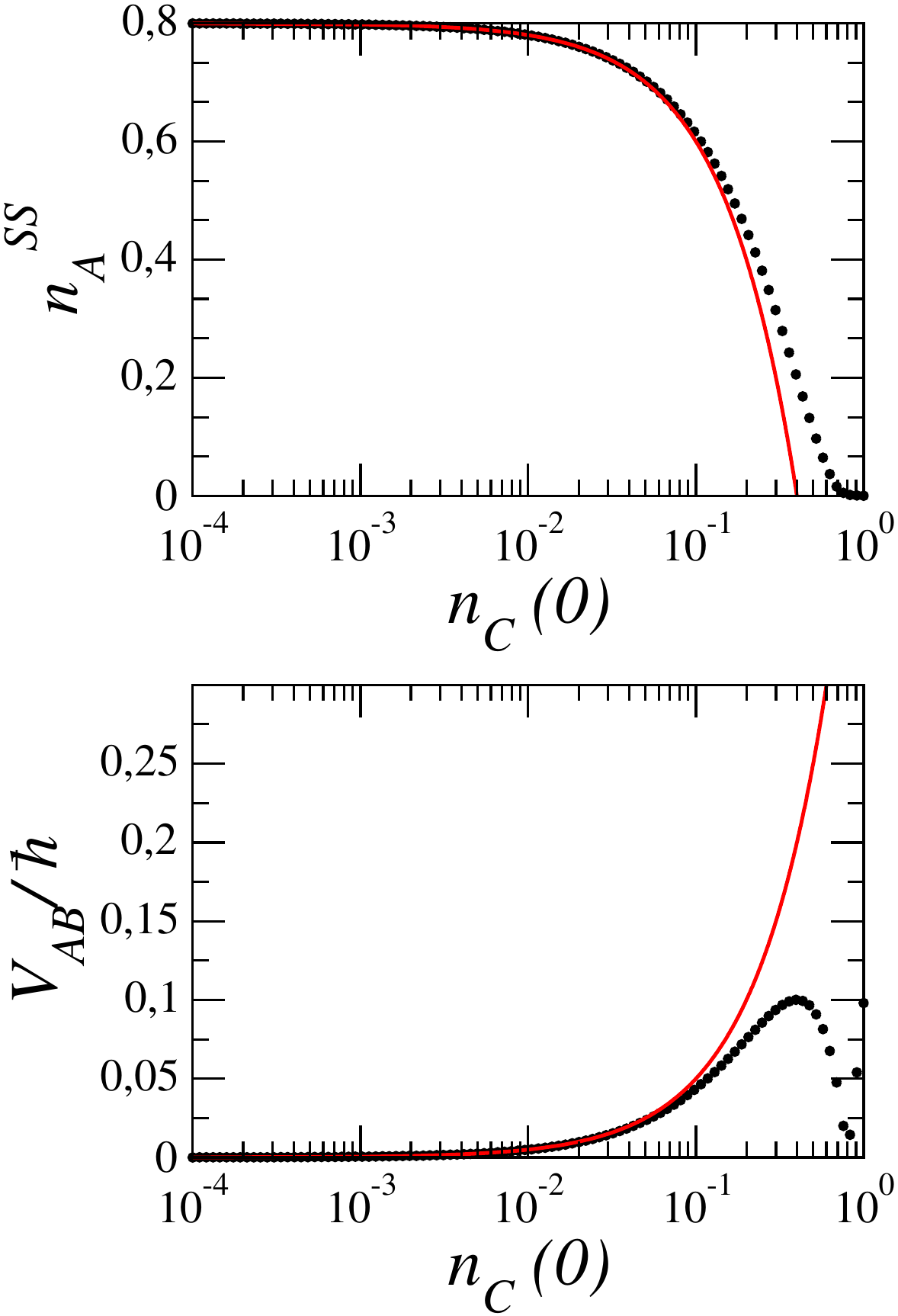}
\caption{SU($3$) residual population (top panel) and violation of the spin conservation (bottom panel). 
Here we set $n_{A}(0)=0.8$ while $n_{B}(0)=n_{C}(0)$ is varied. The numerics shows a good agreement with the prediction of Eq.~\eqref{su3limA} and Eq.~\eqref{su3limAbis}.}
\label{Fig:SU3_sz}
\end{figure}
In the long time limit the system gets empty in the $B,C$ subspaces and display a non-vanishing density in the $A$ sector  
\begin{equation}
\label{su3limA}
n_{A}^{\rm SS} \doteqdot \lim_{t\to\infty}n_{A}(t) = n_{A}(0)-2 n_{B}(0), \qquad \lim_{t\to\infty}n_{B,C}(t) = 0.
\end{equation}
The result~\eqref{su3limA} also allow us to compute the violation of the spin conservation in the $AB$ sectors, in the steady-state we get 
\begin{equation}
\label{su3limAbis}
V_{AB} = s_{AB}(0)-s_{AB}(\infty)= \frac{\hbar}{2} n_{B,C}(0).
\end{equation}
The agreement with the numerical simulation is good and it is shown in Fig.~\ref{Fig:SU3_sz}.

\section{VI. Derivation of the effective master equation in the strongly dissipative and interacting quantum Zeno regime}
\label{App:deriv:eff:me}
In this Section we detail the derivation of the effective master equation governing the dynamics in the Zeno regime in the main text, following the method employed in Ref.~\cite{GarciaRipoll_2009}. As anticipated, the case study is when $\hbar \gamma \gg J$, for this reason it will be useful to regroup the terms of the master equation in the following manner:
\begin{subequations}
\begin{align}
    \frac{d\rho}{dt}&=(\mathcal{V} +\mathcal{L}_{int}) \rho,\\ 
    \mathcal{V} [\rho]&= - \frac{i}{\hbar} [H_J, \rho], \\
    \mathcal{L}_{int} [\rho]&= - \frac{i}{\hbar} [H_{int}, \rho]+ \frac{1}{\hbar} \mathcal{D}[\rho],
\end{align}
\label{eq:fme}
\end{subequations}
where:
\begin{subequations}
\begin{align}
H_J & = -J \sum_j \sum_{\sigma = -I}^I \left(c_{j, \sigma}^{\dagger} c_{j+1, \sigma} + \text{H.c.} \right) \\
H_{int} & = U \sum_j \sum_{\sigma < \sigma'} n_{j, \sigma} n_{j, \sigma'} \\
\mathcal{D}[\rho] & = \sum_\alpha L_\alpha \rho L_\alpha^\dagger - \frac{1}{2} \left\{ L_\alpha^\dagger L_\alpha, \rho \right\}
\end{align}
\end{subequations}

This way of rewriting the master equation helps to realize the order of magnitude of the various term: $\mathcal{V}$ is of order $J$, whereas it has been assumed $ \frac{\hbar \gamma}{U}\sim \mathcal{O}(1)$. It should be now more clear that it is possible to tackle the problem by means of a perturbative approach, where $\mathcal{V}$ is considered a perturbation with respect to  $\mathcal{L}_{int}$. 

\subsection{Properties of $\mathcal L_{int}$}

Before entering in the details of the quasi-degenerate perturbation theory procedure, let us focus on the properties of $\mathcal L_{int}$. By means of a generalized version of Kato's method it is possible to expand:
\begin{equation}
    \centering
    \mathcal{L}_{int}= \sum_i \lambda_i \mathcal{P}_i,
    \label{feq3}
\end{equation}
using a complete set of pseudo-projector operators $\{\mathcal{P} \}_i$ with the following properties:
\begin{equation}
    \centering
    \mathcal{P}_i \mathcal{P}_j= \delta_{ij} \mathcal{P}_i, \hspace{0.2cm} \sum_i \mathcal{P}_i=1.
    \label{feq11}
\end{equation}
One can construct the projector operators starting from the right and left eigenvectors of $\mathcal{L}_{int}$. Consequently, one can decompose the density matrix as a sum of contributions coming from different decoupled subspaces:
\begin{equation}
\rho(t)= \sum_i \rho_i (t),
\end{equation}
where $\rho_i$ is the contribution related to the subsapce with $i$ doubly occupied site. All the subspaces but $\rho_0$ decay with a rate which is bigger the higher the number of double occupancy is.

When $J \ne 0$ a coupling between $\rho_0$ and the decaying states is established, this makes the initially stable state dissipative. On the other hand, the rates of the rapid decaying eigenstates will be slightly modified by $\mathcal{V}$, but they are still of $\mathcal{O}(\gamma)$. This causes another important observation, whatever the initial state is, after a short transient $ t \sim \frac{1}{\gamma} $, most of the states will be captured by the eigenstate with the lowest decay rate. For these reasons it is reasonable to set all contributions $\rho_{n \ge 2} \simeq 0$, i.e. restrict the analysis to the subspaces in which there is at most one double occupancy in the whole lattice. Consequently, we are going to focus only on $\rho_0$ and  $\rho_1$. Actually, we will have $\rho_{1a}$ and $\rho_{1b}$ depending on wheter the double occupancy acts on the right or on the left of the density matrix, respectively.

\subsection{Projector operators and eigenvalues}

Let us now construct the projector operators previously introduced and compute the eigenvalues related to the three subsapces of interest $\rho_0$, $\rho_{1a}$ and $\rho_{1b}$. The procedure is simplified by the fact that $ \mathcal{L}_{int}= \sum_{i=1}^L \mathcal{L}_{loc,i} $ is a sum of commuting local operators $\mathcal{L}_{loc,i}$, one for each lattice site:
\begin{equation}
    \centering
    \mathcal{L}_{loc} [\rho] = -\frac{i}{\hbar} \left [\sum_{\sigma < \sigma'} \hat{n}_{\sigma } \hat{n}_{\sigma' }, \rho \right]+ \frac{\gamma}{2} \sum_{\sigma < \sigma'} \big [ c_{\sigma} c_{\sigma'} \rho c_{\sigma' }^{\dagger} c_{\sigma }^{\dagger} - \frac{1}{2} \big (\hat{n}_{\sigma }  \hat{n}_{\sigma' } \rho + \rho \hat{n}_{\sigma }  \hat{n}_{\sigma' } \big ) \big] 
\end{equation}
It is then possible to diagonalize $\mathcal{L}_{int}$ by introducing the basis of projectors:
\begin{equation}
    \centering
     \lvert \sigma,\sigma')= \lvert \sigma\rangle \langle \sigma' \rvert,
     \label{feq5}
\end{equation}
with $\ket{\sigma}$ one of the following vectors: $\ket{0}, \ket{-I}, \dots, \ket{I}, \ket{\sigma, \sigma'}$ for $\sigma < \sigma'$. This is possible thanks to the approximation that was done before, i.e. $\rho_{2,3, \dots}=0$ that reduces the Hilbert space in which $\mathcal{L}_{loc,i}$ acts to the one with occupation numbers smaller or equal to two. The initial states will belong to the subspace $\rho_0$ and given the fact that $\rho_{2,3,\dots =0}$ an at most double occupancy per site is obtained.

 The scalar product is then defined by introducing the adjoint basis $(\sigma',\tau'|$ in such a way that the Frobenius scalar product obeys the following rule: $(\sigma',\tau'|\sigma,\tau)= \delta_{\sigma \sigma'} \delta_{\tau \tau'}$. In this basis $\mathcal{L}_{loc}$ becomes a bidiagonal non symmetric operator:

\begin{multline}
    \mathcal{L}_{loc} |\sigma,\tau)=-\frac{i U}{\hbar} \big (n_{\sigma } n_{\tau } |\sigma,\tau)-|\sigma,\tau) m_{\sigma } m_{\tau } \big )+\\ 
    +\gamma \big [  |n_{\sigma }-1 \hspace{0.1cm} n_{\tau } -1, m_{\sigma }-1 \hspace{0.1cm} m_{\tau }-1) -\frac{1}{2} \big ( \hspace{0.1cm} n_{\sigma } n_{\tau } |\sigma,\tau)+ |\sigma,\tau) \hspace{0.1cm} m_{\sigma } m_{\tau } \big) \big ]
\end{multline}
The kernel of this operator is associated to the subspace of no double occupancy density matrices that can be formally written as:
\begin{equation}
    \centering
    \rho_0= Q_0 \rho Q_0,
\end{equation}
where:
\begin{equation}
    \centering
    Q_0= q_0^{ \otimes L}, \hspace{0.5cm} q_0= \big (\ket{0} \bra{0} + \sum_{\sigma = -I}^I \ket{\sigma} \bra{\sigma}  \big),
\end{equation}
with L being the number of sites. The other subspace of interest is the one with a single double occupancy. There are two set of states of this kind, depending on whether the double occupation projector is acting on the left or on the right of the density matrix:
\begin{equation}
    \mathcal{P}_{1a} \rho = Q_1 \rho Q_0
\end{equation}
\begin{equation}
    \mathcal{P}_{1b} \rho = Q_0 \rho Q_1,
\end{equation}
and:
\begin{equation}
    Q_1= \sum_{i=1}^L  \sum_{\sigma < \sigma'} q_0^{\otimes i-1} \otimes \lvert \sigma\sigma' \rangle \langle \sigma\sigma' \rvert \otimes q_0^{\otimes L-k}. 
\end{equation}

$\mathcal{L}_{loc}$ can be diagonalized in the basis~\eqref{feq5}; we can then have access to both right and left eigenvector, satisfying:
\begin{equation}
    \centering
    (\mathcal{L}_{loc} -\lambda_n) |v_n)=0
\end{equation}
\begin{equation}
    \centering
    (w_n| (\lambda_n-\mathcal{L}_{loc})=0,
\end{equation}
with $ (w_n|v_m)= \delta_{nm} $. The pseudo-projector local operators are then constructed starting from the right/left eigenvectors :
\begin{equation}
    \centering
    \mathcal{P}_n^{loc}= |v_n) (w_n|.
\end{equation}
Diagonalizing $\mathcal{L}_{loc}$ one gets: \\
\begin{equation}
    \centering
    \mathcal{P}_0^{loc}= \sum_{\sigma < \sigma'} |0,0) (\sigma \sigma', \sigma \sigma'| + \sum_{\sigma, \sigma'= -I, \dots, 0, \dots I} |\sigma ,\sigma') (\sigma ,\sigma'|,
\end{equation}
\begin{equation}
    \centering
    \mathcal{P}_{1a}^{loc}= \sum_{\sigma < \sigma'} \sum_{\tau = -I }^I |\sigma \sigma', \tau) (\sigma \sigma' ,\tau|,
\end{equation}
\begin{equation}
    \centering
    \mathcal{P}_{1b}^{loc}= \sum_{\sigma < \sigma'} \sum_{\tau = -I}^I |\tau,\sigma \sigma') (\tau,\sigma \sigma'|,
\end{equation}
\begin{equation}
    \centering
    \mathcal{P}_2^{loc}= \sum_{\sigma < \sigma'} |\sigma \sigma',\sigma \sigma')(\sigma \sigma',\sigma \sigma'|-|0,0)(\sigma \sigma',\sigma \sigma'|
\end{equation}
and corresponding eigenvalues: 
\begin{equation}
    \centering
    \lambda_0=0,
\end{equation}
\begin{equation}
    \centering
    \lambda_{1a}=-\frac{i U}{\hbar}-\frac{\gamma}{2},
\end{equation}
\begin{equation}
    \centering
    \lambda_{1b}= \lambda_{1a}^{*},
\end{equation}
\begin{equation}
    \centering
    \lambda_{2}= -\gamma.
\end{equation}

Now that an expression for the local pseudo-projector operators has been obtained, it is possible to construct the total ones by a linear combination of them in the following way:
\begin{equation}
    \centering
    \mathcal{P}_0= \mathcal{P}_{0}^{loc} \otimes \dots \otimes \mathcal{P}_{0}^{loc},
\end{equation}
\begin{equation}
    \centering
    \mathcal{P}_{1a}= \sum_{i=0}^{L-1} (\mathcal{P}_{0}^{loc})^{\otimes m} \otimes \mathcal{P}_{1a}^{loc} \otimes  (\mathcal{P}_{0}^{loc})^{\otimes L-i-1}, 
    \label{feq6}
\end{equation}
\begin{equation}
    \centering
    \mathcal{P}_{1b}= \sum_{i=0}^{L-1} (\mathcal{P}_{0}^{loc})^{\otimes m} \otimes \mathcal{P}_{1b}^{loc} \otimes  (\mathcal{P}_{0}^{loc})^{\otimes L-i-1}. 
    \label{feq7}
\end{equation}
Each term in the expressions ~\eqref{feq6} and ~\eqref{feq7} contains only a single localized excitation on a given lattice site, then the total one includes all possible linear combination of them.

The action of these pseudo-projector operators on the density matrix reads:
\begin{equation}
    \centering
    \mathcal{P}_0 \rho= \rho_0= Q_0 \rho Q_0 + \frac{1}{2} \sum_i \sum_{\sigma < \sigma'} c_{i \sigma} c_{i \sigma'} Q_1 \rho Q_1 c_{i \sigma}^{\dagger} c_{i \sigma'}^{\dagger},
    \label{feq8}
\end{equation}
\begin{equation}
    \centering
    \mathcal{P}_{1a} \rho= \rho_{1a}= Q_1 \rho Q_0,
    \label{feq9}
\end{equation}
\begin{equation}
    \centering
    \mathcal{P}_{1b} \rho = \rho_{1b}= Q_0 \rho Q_1.
    \label{feq10}
\end{equation}
The Eqs.~\eqref{feq9} and~\eqref{feq10} are simply the application on the right/left (resp.) of the projector $Q_1$. Conversely, Eq.~\eqref{feq8} contains not only the application of $Q_0$ on both sides. An extra term which first projects the density matrix to a double occupancy on both sides of $\rho$ is present, then the fermionic field annihilation/creation operators restore the zero double occupation.

It is possible to show that, up to second order, the master equation related to the zero double occupancy subspace, i.e. $\rho_0$, is reduced to the following one~\cite{GarciaRipoll_2009}:
\begin{subequations}
\begin{align}
\frac{d}{dt} \rho_0 &= \left(\mathcal L_1+ \mathcal L_2 \right) \rho_0 \\
\mathcal L_1 &= \mathcal{P}_{0} \mathcal{L}_{int} \mathcal{P}_{0} \\
\mathcal L_2 &= \sum_{c \in \{1a,1b \}} -\frac{1}{\lambda_c} \mathcal{P}_0 \mathcal{V} \mathcal{P}_c \mathcal{V} \mathcal{P}_0.
\end{align}
\end{subequations}
\subsection{First-order corrections: hard-core fermions}
Let's start analyze the first term of the effective model given by $\mathcal{L}_1$:
\begin{equation}
    \centering
    \mathcal{L}_1 [\rho_0] = \mathcal{P}_{0} \mathcal{V} \mathcal{P}_{0} \rho_0,
\end{equation}
then, exploiting equation~\eqref{feq8}, and the fact that when one tries to project $Q_1$ on $\rho_0$ the result is zero, one gets:
\begin{equation}
    \centering
    \mathcal{L}_1 [\rho_0]= Q_0 \bigg (-\frac{i}{\hbar} \bigg ) \big [H_J, Q_0 \rho_0 Q_0   \big] Q_0= - \bigg(\frac{i}{\hbar} \bigg ) \big [Q_0 H_J Q_0, \rho_0 \big ],
    \label{feq16}
\end{equation}
where the last equality holds due to the fact that $Q_0$ and $\rho_0$ commute. The physical interpretation of equation ~\eqref{feq16} is that the first order Liouville operator is equivalent to a Hamiltonian that has been projected within states without double occupancies. This is precisely a hard-core fermion gas described by the following master equation (to lowest order):
\begin{subequations}
\begin{align}
     \frac{d \rho_0}{dt} &= -\frac{i}{\hbar} \bigg[H_1, \hspace{0.05cm}  \rho_0   \bigg] + \mathcal{O} (J^2/ |U|) \\
     H_1&= -J \sum_{i=1}^L \sum_{\sigma = -I}^I \big (f_{i+1 \sigma}^{\dagger} f_{i \sigma} + \text{H.c.} \big)
\end{align}
\end{subequations}
where $f_{i \sigma}^{\dagger}$ and $f_{i \sigma}$ are the hard-core fermionic operators satisfying the Clifford algebra plus the hard-core constraint:
\begin{equation}
    \centering
    f_{i \sigma}=\lvert 0 \rangle_i \langle \sigma\rvert_i, \hspace{0.1cm} f_{i \sigma}^{\dagger}=\lvert \sigma \rangle_i \langle 0\rvert_i, \hspace{0.1cm} \sigma \in \{-I, \dots, I  \}.
\end{equation}
The main result so far is that two body losses in the strong dissipative regime lead to a coherent dynamics given by an hard-core fermion Hamiltonian. 

\subsection{Second-order corrections}
The second order Liouville operator reads:
\begin{equation}
    \centering
     \mathcal{L}_2 = \sum_{c \in \{1a,1b \}} -\frac{1}{\lambda_c} \mathcal{P}_0 \mathcal{V} \mathcal{P}_c \mathcal{V} \mathcal{P}_0.
\end{equation}
Expanding this expression one finds:
\begin{equation}
    \centering
    \mathcal{L}_2 \rho_0=\frac{1}{\lambda_{1a} \hbar^2} \mathcal{P}_0 \big [H_J, Q_1 \big[H_J, \rho_0  \big] Q_0 \big]+ \frac{1}{\lambda_{1b} \hbar^2} \mathcal{P}_0 \big [H_J, Q_0 \big[H_J, \rho_0  \big] Q_1 \big].
\end{equation}
Developing the previous equation one should note that term like $Q_1 Q_0$ are zero since it is not possible to project at the same time onto two different subspaces. This joints to the property $\rho_0=Q_0 \rho_0 Q_0$ gives:
\begin{equation}
    \centering
    \mathcal{L}_2 \rho_0= \frac{\mathcal{P}_0}{\hbar^2} \bigg [ \frac{1}{\lambda_{1a}} \big(H_J Q_1 H_J Q_0 \rho_0- Q_1 H_J Q_0 \rho_0 Q_0 H_J \big) + \frac{1}{\lambda_{1a}^*} \big( Q_0 \rho_0 Q_0 H_J Q_1 H_J- H_J Q_0 \rho_0 Q_0 H_J Q_1\big)\bigg].
\end{equation}
The final projection with $\mathcal{P}_0$ is made following equation~\eqref{feq8}; this gives two kind of terms: the first keeps the ones proportional to $Q_0 H_J Q_1 H_J Q_0$ (due to the first part when is projected both on the left and right on $Q_0$), then it acts on terms that have a double occupancy on both side of the density matrix. Let us introduce $T=Q_1 H_J Q_0 / (-J)$, then:
\begin{equation}
    \centering
    \mathcal{L}_2 \rho_0= \frac{J^2}{\hbar^2} \bigg(\frac{1}{\lambda_{1a}} T^{\dagger} T \rho_0+\frac{1}{\lambda_{1a}^*}\rho_0 T T^{\dagger}   \bigg)-\frac{2J^2}{\hbar^2} \bigg(Re \frac{1}{\lambda_{1a}} \bigg) \frac{1}{2} \sum_i c_{i \uparrow} c_{i \downarrow} T \rho_0 T^{\dagger} c_{i \uparrow}^{\dagger} c_{i \downarrow}^{\dagger}.
\end{equation}

It is now possible to rewrite everything in terms of the hard-core fermion operators. Firstly, we compute the quantity $c_{i, \sigma} c_{i, \sigma'} T$. Let us introduce the orthonormal basis $\ket{\psi}$ for the subspace without double occupancies in the lattice, then:
\begin{equation}
Q_0 = \sum_{\psi} \ket{\psi} \bra{\psi}
\end{equation}
In addition, we consider another orthonormal basis $\ket{\phi_i}$ that accounts a double occupancy on the i-th site, one can then write:
\begin{equation}
Q_1 = \sum_i \sum_{\phi_i} \ket{\phi_i} \bra{\phi_i}= \sum_i Q_{1,i}.
\end{equation} 
Moreover, we have that the following holds:
\begin{equation}
c_{i, \sigma} c_{i, \sigma'} Q1 H_J Q_0 = c_{i, \sigma} c_{i, \sigma'} Q_{1, i} H_J Q_0 =  c_{i, \sigma} c_{i, \sigma'} H_J Q_0.
\end{equation}
The above chain of equalities is due to the fact that the only terms which is not killed by $c_{i, \sigma} c_{i, \sigma'}$ is the one with a double occupancy on the $i$-th site. Consequently, it is not anymore necessary to project over the states $\ket{\phi_i}$. In order to proceed with the calculations, we need to understand what are the matrix elements of $H_J$ that are not killed by a double occupancy on the $i$-th site. Firstly, we must have already a fermion in $i$, either $\ket{\sigma}$ or $\ket{\sigma'}$. Then, we can have the hopping from a neighbor site, i.e. $i-1$ or $i+1$. In formulas:
\begin{equation}
c_{i, \sigma} c_{i, \sigma'} H_J Q_0 = \sum_{\psi} c_{i, \sigma} c_{i, \sigma'} \left[ \underbrace{c_{i, \sigma'}^{\dagger} \left( c_{i-1, \sigma'} + c_{i-1, \sigma'} \right)}_{\text{there was a $\sigma$ in i}} + \underbrace{c_{i, \sigma} \left(c_{i-1, \sigma}+ c_{i-1, \sigma} \right)}_{\text{there was a $\sigma'$ in i}} \right] \ket{\psi} \bra{\psi}.
\end{equation}
Eventually, exploting the anticommutation relations and by projecting on $Q_0$ (i.e. take into account the hardocore constraint) we are left with:
\begin{equation}
c_{i, \sigma} c_{i, \sigma'} H_J Q_0 = \left[  \left( f_{i, \sigma} f_{i+1,\sigma'} - f_{i, \sigma'} f_{i+1,\sigma} \right) + \left( f_{i, \sigma} f_{i-1,\sigma'} - f_{i, \sigma'} f_{i-1,\sigma} \right) \right]
\end{equation}
The new quantum jump operators describing the lossy dynamics, characterized by the rate $\gamma_{\text{eff}}$, are thus given by:
\begin{equation}
    \centering
    L_{i}= \left[  \left( f_{i, \sigma} f_{i+1,\sigma'} - f_{i, \sigma'} f_{i+1,\sigma'} \right) + \left( f_{i, \sigma} f_{i-1,\sigma'} - f_{i, \sigma'} f_{i-1,\sigma} \right) \right]
\label{eq:jumps}
\end{equation}
Hence, the new lossy processes consist in the annihilation of a singlet state concerning two nearest-neighbor sites in the $\sigma \sigma'$ subspace. The long-time dynamics is confined in the zero double occupancy subspace whose decay rate $\Gamma_{\text{eff}}$ scales as $\sim \frac{J^2}{\hbar^2 \gamma}$. This is, again, a typical result of the many-body Quantum Zeno effect. 

Summarizing, the second order corrections can be written as:
\begin{equation}
\mathcal{L}_2 [\rho]  =  -\frac{i}{\hbar} \left[ H_2, \rho \right] + \sqrt{\Gamma_{\text{eff}}} \sum_i \left[ L_i \rho L_i^{\dagger} - \frac{1}{2} \left \{L_i^{\dagger} L_i, \rho  \right \} \right], 
\end{equation}
with
\begin{subequations}
\begin{align}
\centering
H_2 = & -J_2 \sum_{i} L_{i}^{\dagger} L_{i},\quad J_2 = \frac{ 2 J^2}{\hbar^2} \Im\left({\frac{1}{\lambda_{1a}}}\right)  \\
\Gamma_{\rm eff} = &\frac{ 2 J^2}{\hbar^2} \Re\left({\frac{1}{\lambda_{1a}}}\right).
\end{align}
\end{subequations}
In the main text we have neglected $H_2$ since it does not affect the no-click dynamics which is dominated by $H_1$.
\section{SU(3) dynamics in the $AC$-$BC$ subspaces in the Zeno regime}
\begin{figure}[t]
\includegraphics[width=0.35\columnwidth]{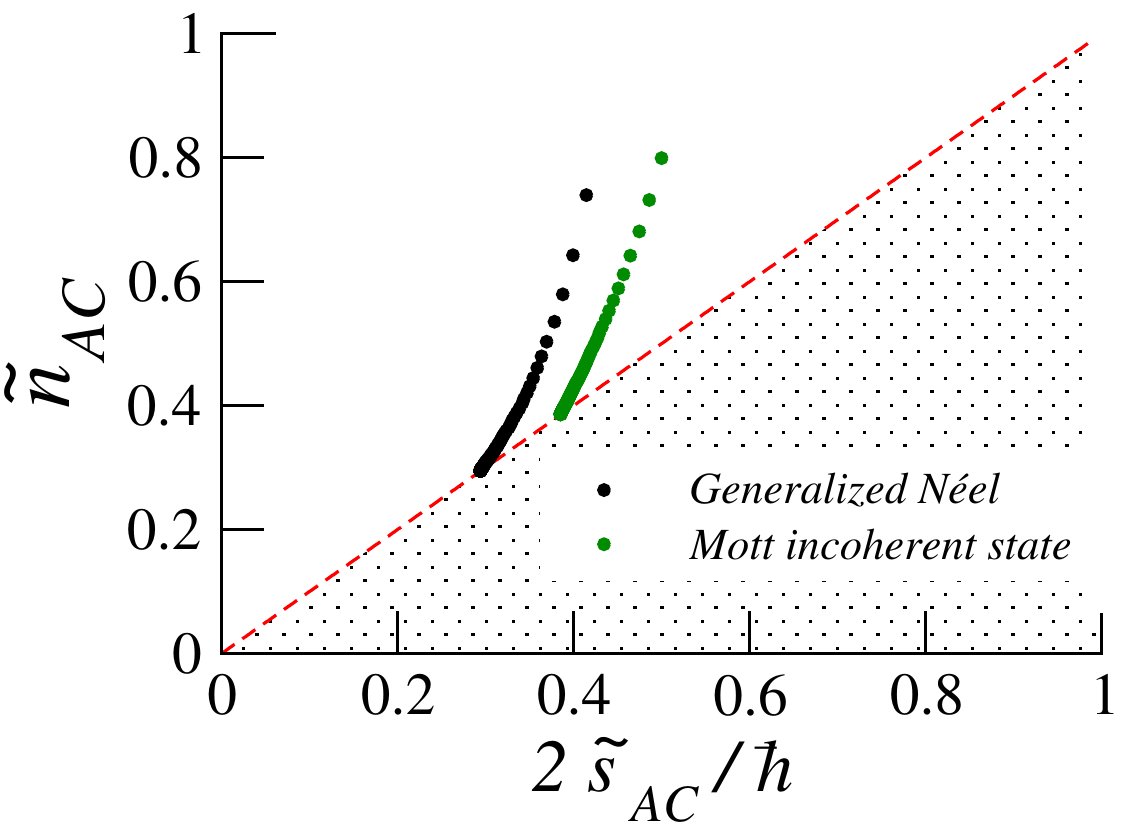}
\includegraphics[width=0.35\columnwidth]{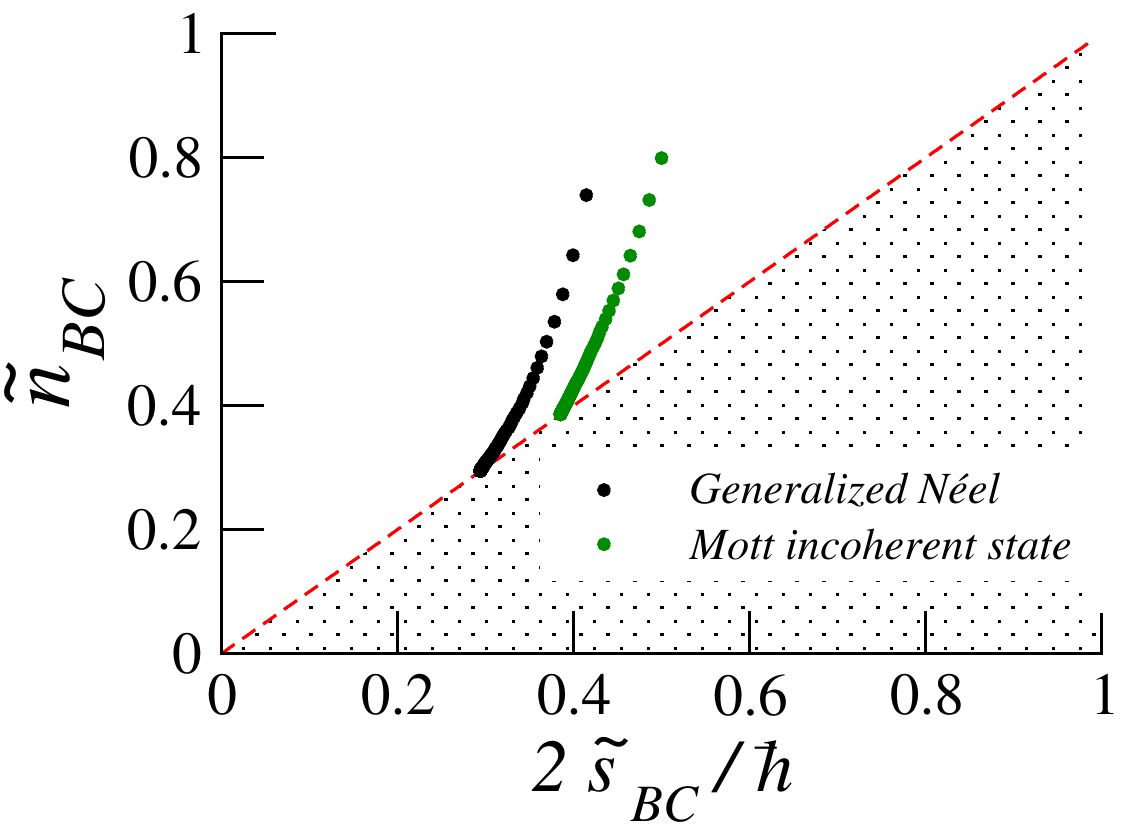}
\caption{SU($3$) dynamics in the $2\tilde{s}_{\mu \mu'}/\hbar - \tilde{n}_{\mu \mu'}$ plane in the $AC$ (left panel) and $BC$ (right panel) subspaces. Orange circles: dynamics from the generalized N\'eel state. Green squares: dyanmics from the Mott incoherent state. The dashed line represents the Dicke cone satisfying Eq.~(6) (main text). Data obtained with $L=8$ and $N_{\rm traj}=2000$}
\label{Fig:SU3:QZ:AB:BC}
\end{figure}

In this section we present the dynamics in the $2\tilde{s}_{\mu \mu'}/\hbar - \tilde{n}_{\mu \mu'}$ for the $AC$ and $BC$ subspaces in the strongly interacting and dissipative limit. In Fig.\ref{Fig:SU3:QZ:AB:BC} we show the dynamics for the other spin subspaces not shown in the main text. As it was already stated in the main text, the dynamics from the Mott incoherent state is independent on the subspace considered, given its rotational invariance. On the other hand, the generalized N\'eel state has different dynamics when considering different subspaces; nonetheless, in this particular case for $L=8$, we found the same dynamics in the $AC$ and $BC$ subspaces given the symmetries of the initial state.
\end{document}